\documentclass[sigconf]{acmart}
\renewcommand\footnotetextcopyrightpermission[1]{} % removes footnote with conference information in first column
\usepackage{tikz}
\usepackage{amsmath}
\usepackage{booktabs} % For formal tables
\usepackage[utf8]{inputenc}
\usepackage{indentfirst}
\usepackage{enumitem}
\usepackage{listings}
\usepackage{multirow}
\usepackage{graphicx}
\usepackage{subcaption}
\usepackage{amssymb}
\usepackage{upgreek}

\def\fixme#1{\typeout{FIXED in page \thepage : {#1}}
\bgroup \color{red}{[FIXME: {#1}]} \egroup}

\setcopyright{none} % No copyright notice required for submissions
\acmConference[Anonymous Submission to ACM CCS 2020]{ACM Conference on Computer and Communications Security}{Due 5 May 2020}{Orlando, TBD}
\acmYear{}

\begin{document}

%don't want date printed
%\date{}

\title{SpectreRewind: Leaking Secrets to Past Instructions}

%for single author (just remove % characters)
% \author{
% {\rm Jacob Fustos}\\
% University of Kansas
% \and
% {\rm Michael Bechtel}\\
% University of Kansas
% copy the following lines to add more authors
% \and
% {\rm Heechul Yun}\\
% University of Kansas
%} % end author

%\maketitle

%\title{SpectreRewind: Strategies for Leaking Secrets to Past Instructions}

  \author{Jacob Fustos}
  \affiliation{%
   \institution{University of Kansas}
 %   \city{Lawrence}
 %   \state{Kansas}
  }
%  \email{jacobfustos@ku.edu}

\author{Michael Bechtel}
  \affiliation{%
   \institution{University of Kansas}
% %   \city{Lawrence}
% %   \state{Kansas}
  }

  \author{Heechul Yun}
  \affiliation{%
   \institution{University of Kansas}
% %   \city{Lawrence}
% %   \state{Kansas}
  }
%  \email{heechul.yun@ku.edu}

% \renewcommand{\shortauthors}{}

\begin{abstract}

Transient execution attacks utilize micro-architectural covert channels to leak secrets that should not have been accessible during logical program execution. Commonly used micro-architectural covert channels are those that leave lasting footprints in the micro-architectural state, for example, a cache state change, from which the secret is recovered after the transient execution is completed. 
%This lasting footprints has led attackers to utilize an attack framework where secrets are transmitted into covert channel during transient execution and later, after transient execution is complete, read secret from covert channel.

In this paper, we present SpectreRewind, a new approach to create contention based covert channels for transient execution attacks. 
In our approach, a covert channel is established by issuing the necessary instructions logically \emph{before} the transiently executed victim code. 
Unlike prior contention based covert channels, which require simultaneous multi-threading (SMT), SpectreRewind supports single hardware thread based covert channels, making it viable on systems where attacker cannot utilize SMT. We show that contention on the floating point division unit on commodity processors can be used to create a high-performance ($\sim$100 KB/s), low-noise covert channel for transient execution attacks instead of commonly used flush+reload based cache covert channels.

We implement a Meltdown attack utilizing the proposed covert channel showing competitive performance compared to the state-of-the-art cache based covert channel implementation. We also show that the covert channel works in the JavaScript engine of a Chrome browser.

\end{abstract}
%% \ccsdesc[500]{Computer systems organization~Security}
%% % \ccsdesc[300]{Computer systems organization~Redundancy}
%% % \ccsdesc{Computer systems organization~Robotics}
%% % \ccsdesc[100]{Networks~Network reliability}
%\keywords{Spectre, Micro-architecture, Side-channel Attack}
%\maketitle

\begin{CCSXML}
<ccs2012>
<concept>
<concept_id>10002978.10003006.10003011</concept_id>
<concept_desc>Security and privacy~Browser security</concept_desc>
<concept_significance>500</concept_significance>
</concept>
<concept>
<concept_id>10002978.10003001.10010777.10011702</concept_id>
<concept_desc>Security and privacy~Side-channel analysis and countermeasures</concept_desc>
<concept_significance>500</concept_significance>
</concept>
</ccs2012>
\end{CCSXML}

%\ccsdesc[500]{Security and privacy~Side-channel analysis and countermeasures}

\maketitle

%-----------------------------------------------------------------------
\section{Introduction}

Modern out-of-order microprocessors support speculative execution to improve performance. In speculative execution, 
instructions can be executed speculatively before knowing whether they are in the correct program execution path. 
If the speculation was wrong, the instructions that were executed incorrectly---known as transient instructions~\cite{spectre_v1_v2_2019_Kocher}---are squashed and the processor then simply retries to fetch and execute the correct instruction stream. 
Unfortunately, it turned out that these transient instructions can potentially bypass both software and hardware defenses to access secrets. The disclosure of Spectre~\cite{spectre_v1_v2_2019_Kocher} and Meltdown~\cite{meltdown_2018_lipp} and many other subsequently disclosed transient execution attacks~\cite{spectre_v1.1_1.2_2018_kiriansky,spectre_ret2spec_2018_maisuradze,spectre_ret_stack_2018_koruyeh,spectre_v4_2018_horn, meltdown_3a_arm_2018,meltdown_3a_intel_2018,meltdown_lazy_fp_2018_stecklina,meltdown_foreshadow_2018_vanbulck,meltdown_foreshadow-NG_2018_weisse,mds_ridl_2019_Schaik,mds_fallout_2019_Minkin,mds_zombie_2019_Schwarz, transient_systematic_2018_canella, mds_cacheOut, meltdown_LVI} have shown the danger of these transient instructions, as the secrets they had access to could be encoded and transmitted into microarchitectural covert channels, from which normal, non-speculative instructions could then read, allowing the secrets to be visible to the attacker.

All known transient execution attacks share the same three basic steps: (1) the attacker initiates speculative execution where the secret is read improperly from memory or registers; (2) the secret dependent transient instructions then encode and transmit the secret to a micro-architectural covert channel; (3) finally, the secret is recovered from the covert channel by normal (non-transient) receiver instructions. 
Commonly used covert channels, such as cache, are stateful as they leave lasting footprints in the micro-architectural state, from which the secret is recovered after the transient execution is completed. 
Many hardware defense proposals aim to prevent such stateful covert channels either by  hiding the changes into additional hardware buffers~\cite{solution_safespec_2018_khasawneh, solution_invisispec_2018_yan2018, solution_RISCV_2019_Gonzalez} or by reverting them~\cite{solution_CleanupSpec_2019_Saileshwar} when the transient instructions are squashed. Such a mitigation strategy is attractive from a performance standpoint, as the transient instructions are allowed to execute normally, retaining many of the performance benefits of speculative execution.

These type of defenses are effective at blocking transient execution attacks that utilize stateful covert channels. Unfortunately, these techniques cannot be used to block attacks that both transmit into and read from covert channel before transient instructions have been squashed.
SmotherSpectre~\cite{port_smother_2019_bhattacharyya} is the first to demonstrate such an attack, by utilizing a simultaneous covert channel in a simultaneous multi-threading (SMT) setup where contention on issue ports within the processor is used as a covert channel to transmit secret between the hardware threads in the context of Spectre-based attack. Such contention cannot be buffered or reverted, as instructions have already waited to use the issue ports, affecting their execution time.

% our work: we are first to show the feasibility of non-SMT, non-cache based speculative execution attacks.
In this paper, we present a new class of contention-based covert channels for transient execution attacks, which we call SpectreRewind. Like SmotherSpectre, SpectreRewind allows the attacker to both transmit and receive secret data before transient execution has completed, allowing the attacker to bypass most defense mechanisms that attempt to revert or hide micro-architectural changes caused by the attack. However, unlike SmotherSpectre, SpectreRewind does not require the attacker to utilize SMT, instead the attack can be executed from a single hardware thread. 
While traditional transient attacks locate the instructions that will read from the covert channel logically \emph{after} the instruction that triggers the transient execution (e.g., a branch), SpectreRewind takes the opposite approach and locates these instructions logically \emph{before} the triggering instruction. This structure allows the transmitting and receiving instructions to execute concurrently on a modern out-of-order core and communicate the secret even before the transient execution completes.

We start by presenting our SpectreRewind strategy and examining the unique challenges associated with it (Section ~\ref{sec:DesignCC} and Section~\ref{sec:whypipeline}). 
We then apply SpectreRewind to create a new covert channel, which utilizes contention on a \emph{floating point division unit} in commodity Intel and AMD processors (Section ~\ref{sec:CovertChannel}). 
% Next, we present our SpectreRewindMDS strategy (Section ~\ref{sec:DesignMDS}) and then apply SpectreRewindMDS on load port sampling to create a POC, which is able to leak each value of a linked list loaded from a victim, as well as the return addresses that the victim jumps to (Section ~\ref{sec:MDS_POC}).
Next, we analyze how SpectreRewind can be integrated into different transient execution attacks, modifying the Meltdown POC programs to utilize our covert channel, and implementing our covert channel within the Google Chrome JavaScript sandbox (Section ~\ref{sec:SideChannel}). 
Finally, we discuss the security implications that this attack has on currently proposed and implemented hardware and software defenses (Section ~\ref{sec:eval}). Our evaluation results show that our technique allows the creation of a simultaneous covert channel from a single threaded context that has low noise, and is viable for use in transient execution attacks, and can leak data at a rate up to 100 KB/s with an error rate of less than 0.01\%.

\section{Background}\label{sec:background}

In this section, we provide necessary background on out-of-order cores, transient execution attacks, and simultaneous multithreading (SMT) hardware. 

\subsection{Out-of-order Processors}

\begin{figure}[h]
\centering
  \includegraphics[width=0.45\textwidth]{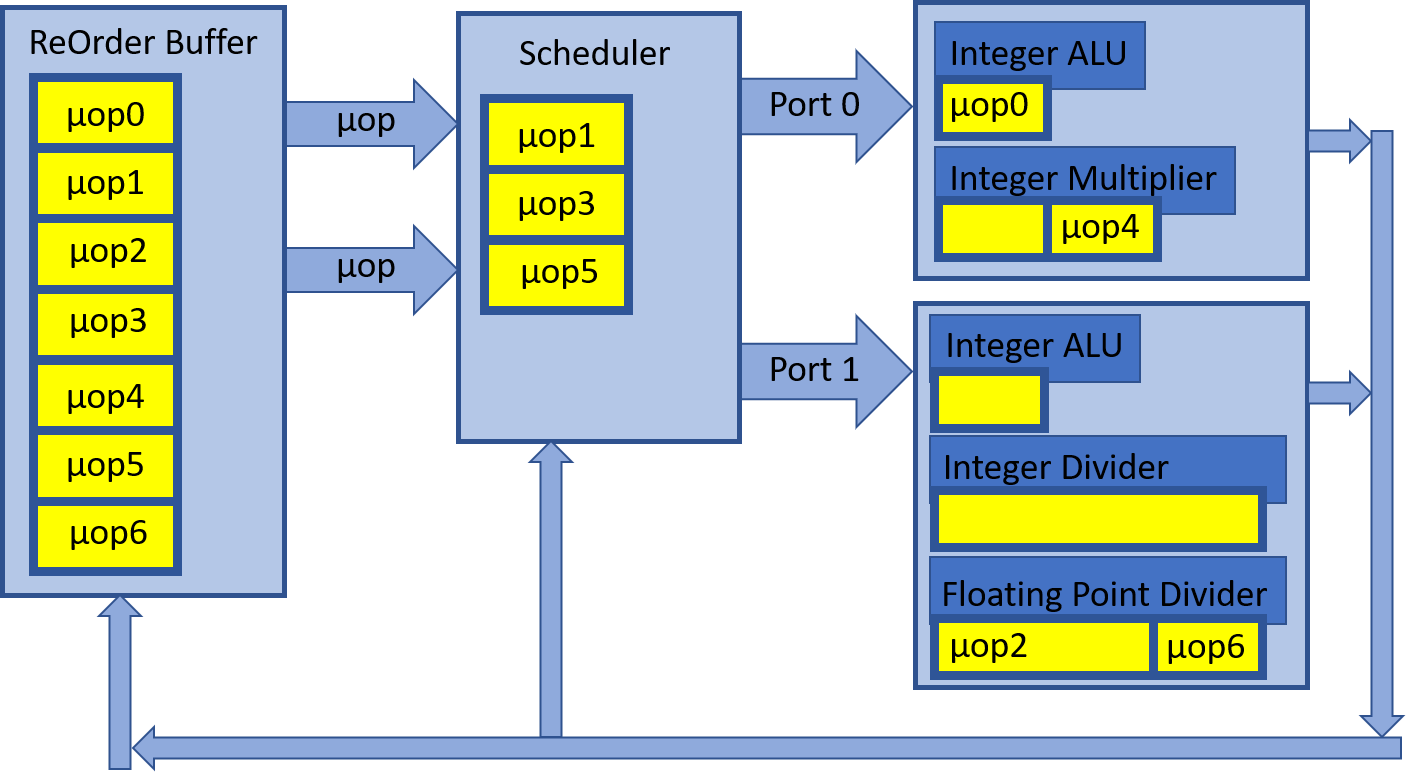}
  \caption{\label{fig:backend} Simplified out-of-order processor design. The ReOrder Buffer holds and retires~$\upmu$ops in logical program order, while~$\upmu$ops are issued to the execution units in out-of-order.}
\end{figure}

Modern high performance microprocessors utilize out-of-order execution to execute multiple independent instructions in parallel---taking advantage of instruction level parallelism---allowing for higher throughput, while also reducing the penalty of a stall caused by independent instructions. 

Figure~\ref{fig:backend} shows a simplified example of an out-of-order processor. 
In this example, instructions have first been translated into micro-operations ($\upmu$ops). 
%This translation can be useful, especially in CISC architectures where the individual steps of complex instructions can be broken into multiple µops extracting further parallelism from single instructions. 
These~$\upmu$ops are first placed into the ReOrder Buffer (ROB) in logical program order. 
They are then passed to the scheduler where they are then issued to a proper functional unit, once their operands become available and the necessary resources are available. In this example, the functional units are clustered into two execution units. Each execution unit contains a single issue port, which can only issue a single~$\upmu$op to one of the enclosed functional units every clock cycle, but once issued, the functional units run independent of each other. Once executed by a functional unit, the scheduler is notified so that it can forward the results to following dependent~$\upmu$ops. The~$\upmu$op then waits in the ROB until it reaches the head where it may be retired. It is only now that the changes made by the~$\upmu$op become architecturally visible, giving the illusion---from the architecture's point of view---that the instructions are executed in-order. 
%Since we only focus on µops and the instructions that we are interested in are translated into a single µop we will use these 2 terms interchangeably.

To further reduce branch related stalls, modern processors implement complex branch predictors to predict what instructions should be executed. As the results of execution are not made architecturally visible until the instructions retire, it is architecturally safe to create a checkpoint of the processor's state before the prediction and then store the predicted instruction stream in the ROB and execute the instructions---speculative execution---further improving performance if the prediction was correct. If the prediction was false, these instructions are squashed---returning the processor state to the checkpoint---where the processor can begin executing the correct instructions. Instructions that were executed but were then squashed---will never become architecturally visible---are known as transient instructions.

\subsection{Transient Execution Attacks}\label{sec:trans_attacks}

As transient instructions were not supposed to have executed, it follows that they can perform tasks---such as accessing secret data---that should not have been accessible during proper program execution. While they do not retire---and do not become architecturally visible---they still can contend for shared resources with instructions that will retire, creating a microarchitectural side-channel that can leak the secrets.

\begin{figure}[h]
\begin{verbatim}
       if (x < array1_size)
               secret = array1[x];
               y = array2[secret * 4096];
\end{verbatim}
\caption{A Spectre gadget. Adopted from~\cite{spectre_v1_v2_2019_Kocher}}
\label{fig:spectregadget}
\end{figure}

Figure~\ref{fig:spectregadget} shows an example of a speculative execution attack---Spectre variant 1~\cite{spectre_v1_v2_2019_Kocher}---in this example, the if statement of line 1 has been trained by the attacker such that the body---lines 2 and 3--- is executed, even though x is out-of-bounds---transient execution. This allows the attacker to chose the value of x such that it points to the value of the secret and when line 2 is executed, the value of the secret---something that would have been blocked during normal program flow---will be unintentionally loaded from memory. The secret value is then encoded in line 3 to leak its value into the cache state. Later---after the transient instructions have been squashed---the attacker can read the cache state---from architecturally visible instructions---and decode it to complete the side-channel.

Transient execution attacks can be broken down into two categories. Spectre type attacks utilize control- and data-flow mis-speculation to force a victim to access secrets from their own address space and leak them into the covert channel where they can be accessed by the attacker. Each Spectre variant---1~\cite{spectre_v1_v2_2019_Kocher}, 1.1~\cite{spectre_v1.1_1.2_2018_kiriansky}, 2~\cite{spectre_v1_v2_2019_Kocher}, 4~\cite{spectre_v4_2018_horn}, and ret2spec~\cite{spectre_ret2spec_2018_maisuradze, spectre_ret_stack_2018_koruyeh}---is distinguished by the microarchitectural component that is responsible for causing the mis-speculation namely---Branch History Buffer (BHB), Branch Target Buffer (BTB), Memory Disambiguator, Return Stack Buffer (RSB). Meltdown style attacks take advantage that processor exceptions are deferred until the instruction that caused the exception is retired---becomes architecturally visible---instructions that occur logically after the exception should not execute in regards to logical program order---but can be executed out-of-order---potentially bypassing the security that the exception intended to provide. These attacks can be run from within the attacker's own address space while allowing them to access secrets from other processes and privilege levels. Each Meltdown variants---1.2~\cite{spectre_v1.1_1.2_2018_kiriansky}, 3~\cite{meltdown_2018_lipp}, 3a~\cite{meltdown_3a_arm_2018, meltdown_3a_intel_2018}, Lazy FP~\cite{meltdown_lazy_fp_2018_stecklina}, and L1TF~\cite{meltdown_foreshadow_2018_vanbulck, meltdown_foreshadow-NG_2018_weisse}---correspond to the exception that caused the fault. Microarchitectural Data Sampling(MDS)~\cite{mds_fallout_2019_Minkin, mds_ridl_2019_Schaik, mds_zombie_2019_Schwarz} are also considered Meltdown-type attacks. These attacks target speculative loads that have incorrectly loaded data from internal buffers---Store Buffer, Load Port, Line Fill Buffer---and leak the data into covert channels before realizing the fault. The data that was incorrectly loaded could have come from other SMT threads on the same processor executing at any privilege level.

\subsection{Simultaneous Multithreading (SMT)}
%Looking back at Figure~\ref{fig:backend}, we see that many of the functional units are not utilized. 
To improve hardware utilization, 
%of costly resources---such as functional units---instead of adding more individual cores to a processor, 
manufacturers often employ a technique called Simultaneous Multithreading (SMT)~\cite{Tullsen-1995-SMT}, where a single core is allowed to execute multiple hardware threads---instruction streams---simultaneously. These hardware threads share the underutilized structures, improving utilization, while appearing---from the architectural point of view---to be independent processing cores.
Because the hardware threads in a SMT capable core share various hardware structures---e.g. issue ports and functional units---these structures can be used to create covert channels and microarchitectural side channels between processes running on the multiple threads within a single core.

\section{Threat Model}\label{sec:threatmodel}

%\begin{figure}[h]
%\begin{verbatim}
%    char get_byte(char * A, int x )
%    {
%       if (x < A.size)
%          return array1[x];
%       return 0;
%    }
%\end{verbatim}
%\caption{Function containing a partial Spectre gadget}
%\label{fig:threat_model_A}
%\end{figure}

% \begin{figure}[h]
% \begin{verbatim}
%     // attacker controlled code
%     secret = transient_execution_attack();
%     // attacker controlled code
% \end{verbatim}
% \caption{Attacker controlled code located around the transient execution attack}
% \label{fig:threat_model_B}
% \end{figure}
%\fixme{need to be revised, or be part of attack sections.}
%We assume that there exists a secret that the attacker would like to leak either through one of the currently known transient execution attacks listed in Section~\ref{sec:trans_attacks} or through a yet to be discovered transient execution attack. We do not make assumptions about the location of the secret, only that it can be targeted by the chosen attack.
%We assume that the victim contains either a known or unknown transient execution vulnerability.
We assume that the attacker has the ability to control some code that executes both logically before and after a transiently execution attack in program order. 
We assume that the attacker would like to construct code that will transmit the secret over a covert channel such that they may read the value of the secret at the architectural level.
We assume that stateful covert channels, such as caches, are not available to the attacker because the platform either does not provide necessary means to control cache state (e.g., clflush) or implements hardware level defense mechanisms that prevent stateful covert channels~\cite{solution_RISCV_2019_Gonzalez, solution_safespec_2018_khasawneh, solution_invisispec_2018_yan2018, solution_CleanupSpec_2019_Saileshwar}.
%We assume that the attacker cannot successfully either create or coordinate with other attacker-controlled threads or processes that are running on the same system.
%We also assume that the program that the secret resides in is correct---i.e., it does not contain any other vulnerabilities either from architectural or micro-architectural perspective that could allow attacker to leak secret.
\section{SpectreRewind}\label{sec:DesignCC}

%In this section, we present the SpectreRewind approach. 
%provide the design behind our attack framework that can be utilized in transient execution attacks, which we call SpectreRewind. 
%We then show how SpectreRewind can be used to create covert channels that require simultaneous contention of resources, such as contention on shared functional unit.

SpectreRewind is an approach to utilize contention-based covert channels for transient execution attacks. It allows the attacker to both transmit into and receive from a covert channel \emph{before} the transient execution phase of the attack is completed.

\begin{figure}[h]
\centering

  \includegraphics[width=0.45\textwidth]{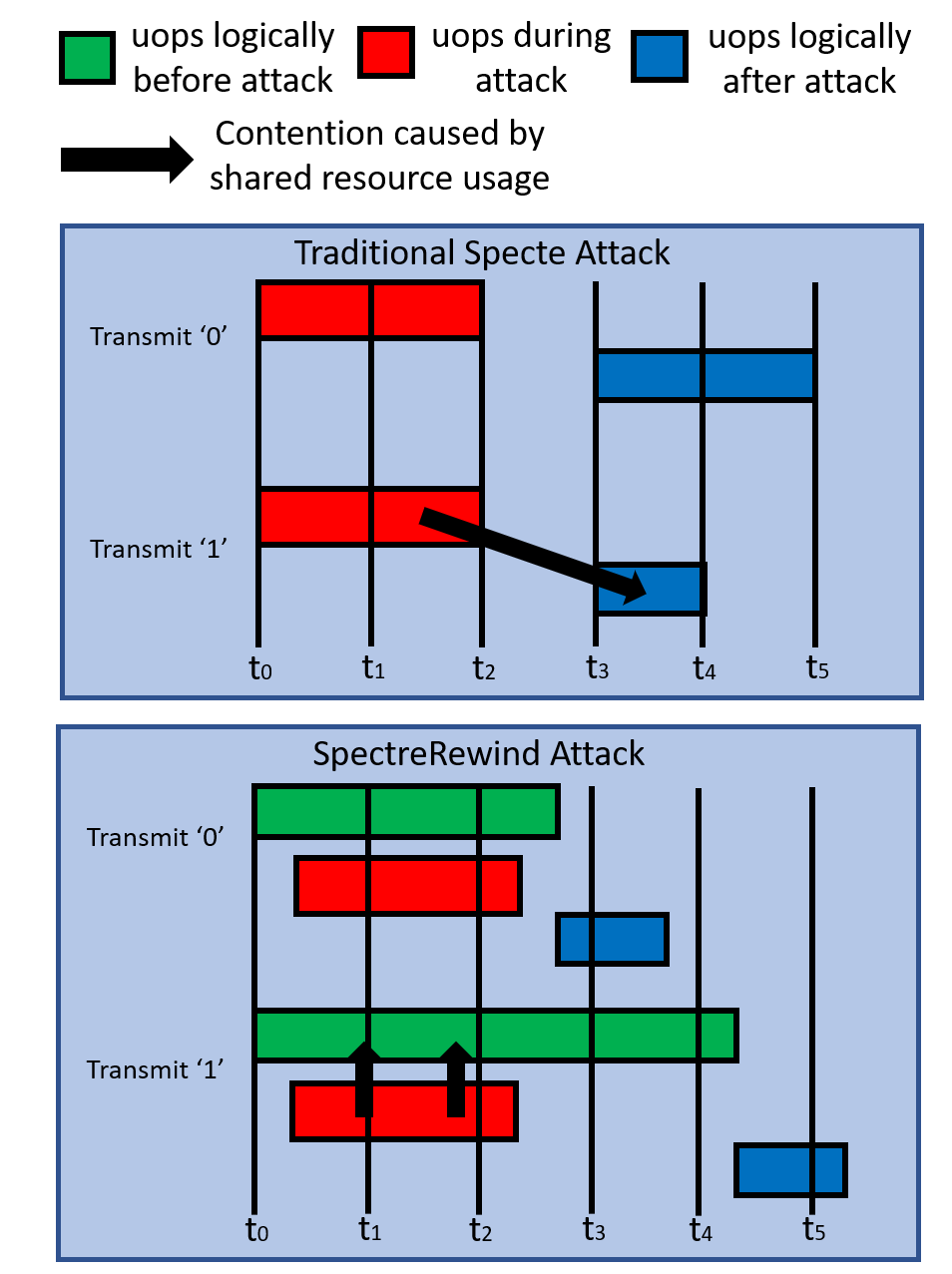}
  \caption{Simplified timing diagram comparing traditional Spectre attack framework to SpectreRewind framework }
  \label{fig:framework}
\end{figure}

% \subsection{SpectreRewind}
%We present our framework for transmitting secrets to µops that come before the transient execution attack even occurs in logical program order. 

Figure~\ref{fig:framework} illustrates the basic concept of SpectreRewind in comparison to the conventional stateful covert channel based approach. 
In this figure, we break up involved~$\upmu$ops into three distinct categories: the~$\upmu$ops that come logically before, during, and after the transient execution attack. In both approaches, we assume that we send only a single bit over a covert channel at a time. For each approach, we depict two timing diagrams: transmitting `0' and `1' over a covert channel. 

In case of the traditional transient execution attack approach, the attacker will use a covert channel that causes a lasting state change in the micro-architecture, and read from the covert channel from~$\upmu$ops that occur logically after the transient execution. 
Data can read from the channel by measuring the timing differences of these~$\upmu$op (t3-t4 for a value `0' and t3-t5 for a value `1'). 
Hardware defenses (e.g., ~\cite{solution_invisispec_2018_yan2018, solution_safespec_2018_khasawneh}) that remove secret from covert channel after transient execution (t2) will be able to stop this attack by disrupting the transmission of the secret.
%allowing the  attempt to transmit the value of the secret by contending with µops that follow the attack in logical program order. As we have shown in the previous section, this will limit the attacker to only be able to use footprint covert channels, but allow the attacker to begin timing to detect either the '0' or '1' value at t3, after the transient instructions have been squashed and no longer occupy space in the ROB. 

In case of the SpectreRewind approach, however, transient instructions will contend for resources with the~$\upmu$ops that come logically before the transient instructions. 
Because the covert channel will be read from before transient execution completes (t2), the aforementioned hardware defense mechanisms which  attempt to remove the secret from the covert channel at that time will be ineffective.
In our approach, the attacker measures the entire execution time of the attack to detect the timing differences. Since the covert channel must be read from before transient execution completes, this gives the added challenge of needing to fit the entire attack in the ROB at the same time.
%which means that both the transient instructions and instructions that will read from the covert channel must exist in the ROB at the same time. This will have the added benefit though of allowing the attacker to utilize simultaneous covert channels to transmit secret values.

%Although any micro-architectural covert channel can be utilized by 
SpectreRewind assumes that older transient~$\upmu$ops can contend with younger~$\upmu$ops that began before the transient~$\upmu$ops on certain micro-architectural resources. In the following, we will discuss the kinds of micro-architectural resources are viable covert channels in SpectreRewind. 

%that they are trying to contend with, this may cause additional challenges that will vary between different micro-architectures. We explore a few of these challenges in the following section.

\section{Not Fully Pipelined Functional Units}\label{sec:whypipeline}

\begin{figure}[htp]
    \centering
    \begin{subfigure}[b]{0.4\textwidth}
        \includegraphics[width=\textwidth]{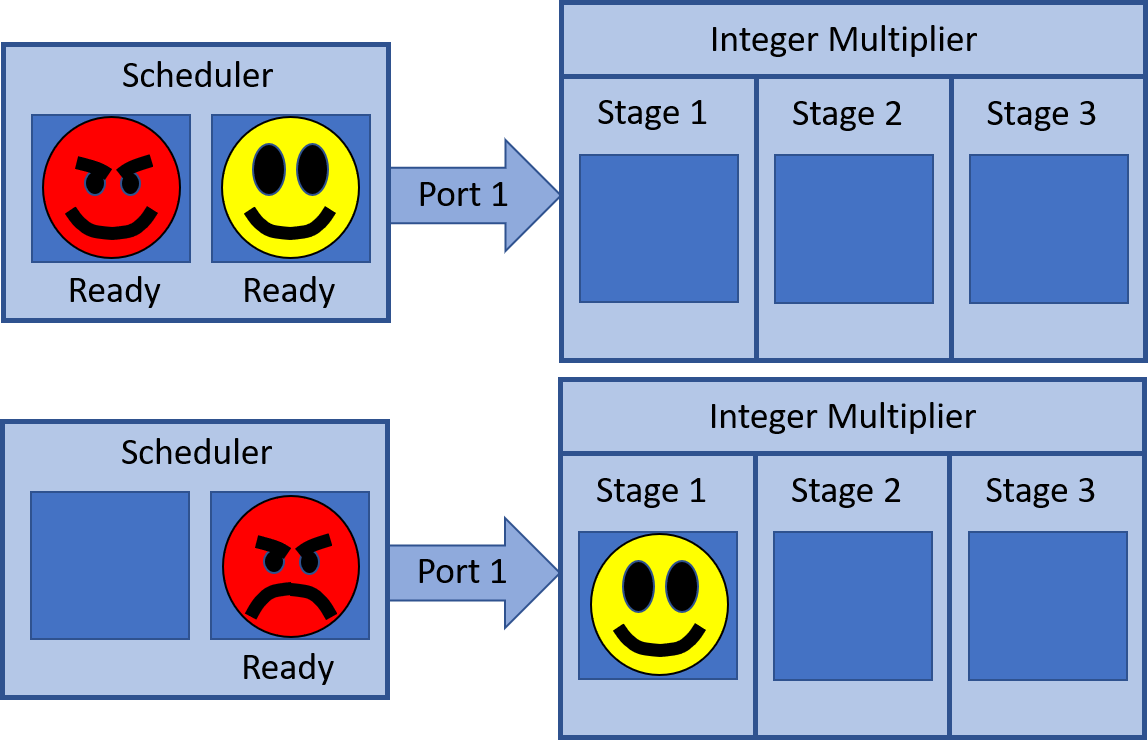}
        \caption{Ready victim, Pieplined functional unit}
        \label{fig:func_try_A}
    \end{subfigure}
    \begin{subfigure}[b]{0.4\textwidth}
        \includegraphics[width=\textwidth]{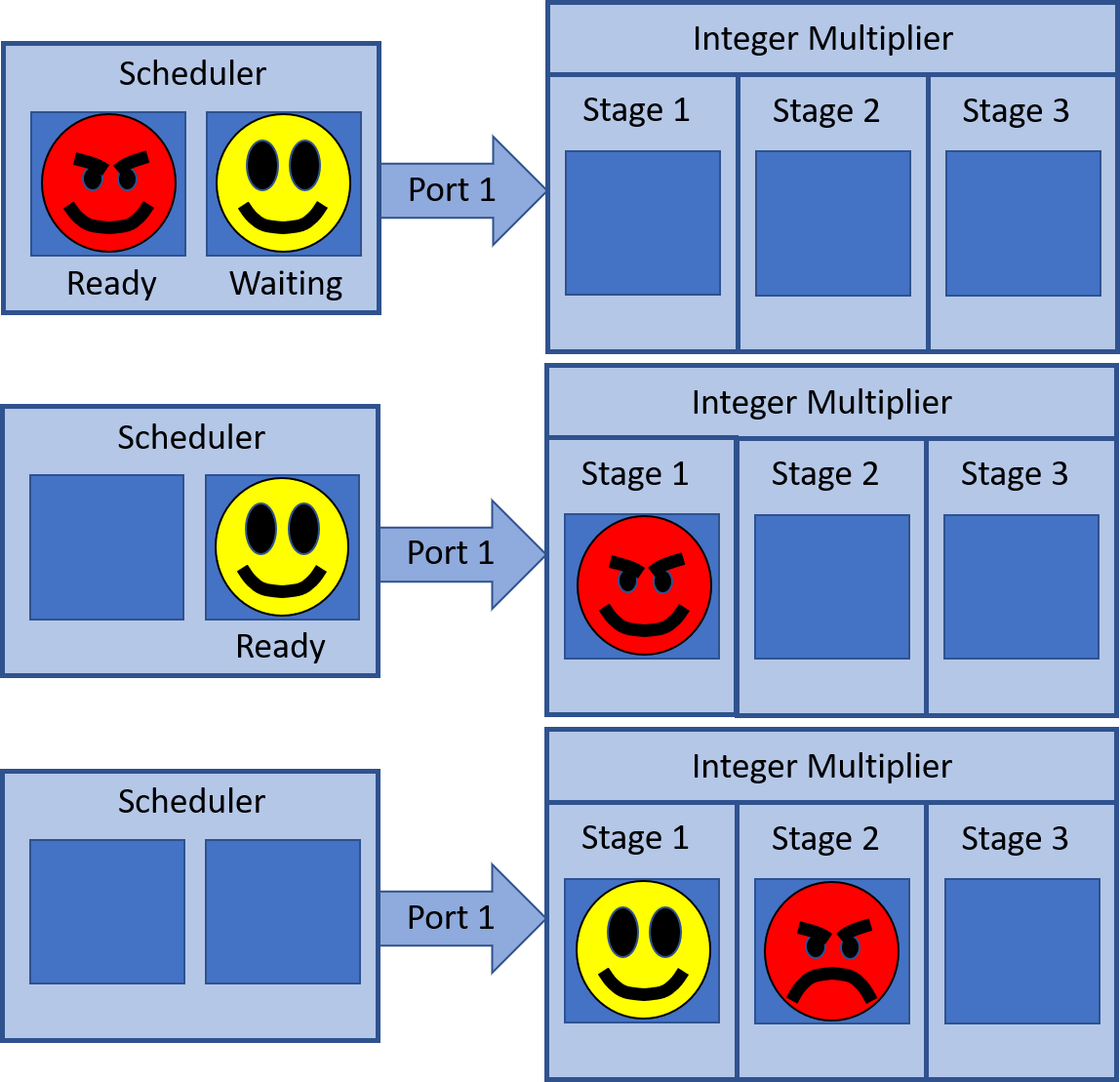}
        \caption{Waiting victim, Pipelined functional unit}
        \label{fig:func_try_B}
    \end{subfigure}
    \begin{subfigure}[b]{0.4\textwidth}
        \includegraphics[width=\textwidth]{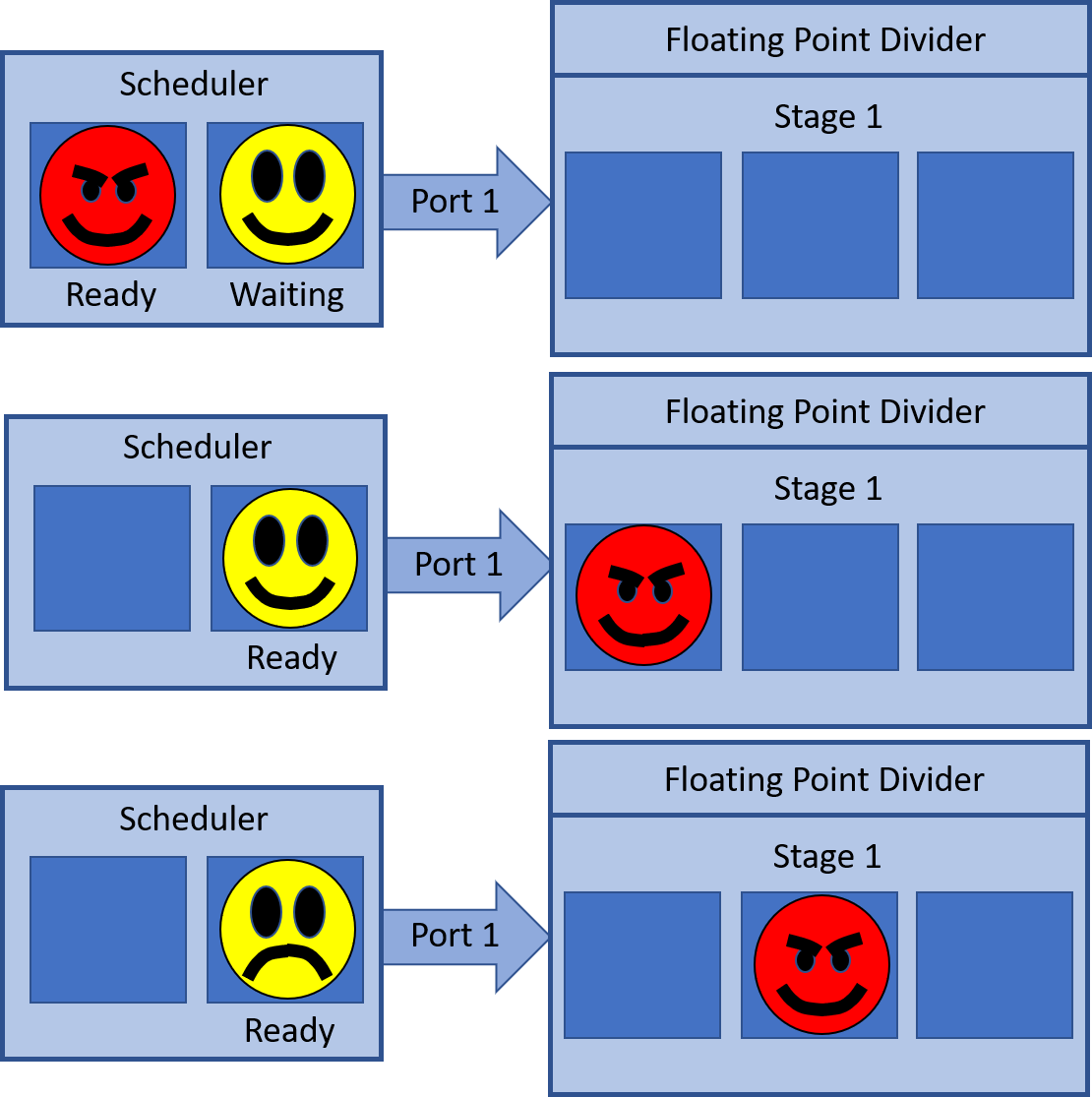}
        \caption{Waiting victim, Not fully pipelined functional unit}
        \label{fig:func_try_C}
    \end{subfigure}
    \caption{ Multiple attempts by attacker to delay the execution of the victim, causing measurable timing differences. If the attacker is younger than the victim, an age-ordered scheduler will prevent most contention.}
    \label{fig:func_try}
\end{figure}

Since we aim to contend with instructions that are logically older than us this will limit our covert channel options. 
We will not be able to cause port contention or contention on fully pipelined functional units as in ~\cite{port_smother_2019_bhattacharyya}. 
However, we will show that it is still possible to cause contention on certain functional units that contain at least one non-pipelined stage.

Figure~\ref{fig:func_try} shows visual examples of this problem. In Figure~\ref{fig:func_try_A}, we see an example of an attacker~$\upmu$op trying to cause slowdown on a victim~$\upmu$op that is trying to use a shared integer multiplier. Unfortunately, because both the attacker and victim are ready to issue, the scheduler will choose the older victim, preventing any contention. 

Figure~\ref{fig:func_try_B}, shows the situation where the attacker becomes ready the cycle before the victim. The attacker is issued into the multiplier, but still cannot create contention on the victim, as the victim is issued on the next cycle that it becomes ready, just as if the attacker was not there. 

Finally, figure~\ref{fig:func_try_C} shows an attack on a non-pipelined shared functional unit (stage 1 takes 3 clock cycles to complete). As the victim is not initially ready, the attacker is scheduled on the unit. As the unit is not pipelined, the victim cannot be issued on the unit until the attacker completes, which effects the execution time of the victim, making a covert channel possible. Thus, for our attack we will only focus on functional units that have at least one stage that is not fully pipelined. 
Note that it is well known that floating point division is difficult to pipeline because for division each step depends on the previous step~\cite{oberman1999floating}. In the following, we will develop a floating point division unit based covert channel. 

% \subsection{Internal Buffers in Non-blocking Caches}
% \fixme{discuss why internal buffers in non-blocking caches such as MSHRs can be used as a SpectreRewind covert channel.}

% \fixme{BTW, MSHRs in a shared cache can be used to create a cross core covert channel.} 
% %\input{Design-example}
\section{Floating Point Division Covert Channel}\label{sec:CovertChannel}

In this section, we utilize our SpectreRewind approach to create a covert channel on real commodity hardware that can transmit data from transient execution without using stateful covert channels, or SMT co-scheduled processes. We do this by causing contention on the floating point division unit.

% \subsection{Transmission Over the Covert Channel}

\begin{figure}[h]
\begin{lstlisting}[linewidth=0.45\textwidth,language=c]
    double recv, div;
    double send1, send2, send3, send4;
    int message; // secret
    
    start = rdtscp(); // start timer
    
    // begin receiver (12 dependent FP divisions)
    recv /= div;        
    recv /= div;        
    ...
    recv /= div;  
    // end of receiver
    
    if (recv == 1) { // begin speculative execution
        m_bit = bit(message, k);  
        if (m_bit) { // secret dependent branch
            // begin sender (independent FP divisions)
            for (int x = 0; x < 100; x++) {
                send1 /= div; 
                send2 /= div; 
                send3 /= div;          
                send4 /= div; 
            }
            // end of sender
        }
    }
    
    end = rdtscp(); // end timer
\end{lstlisting}
\caption{ Pseudo code of our floating point division unit contention based covert channel.}
\label{fig:func_covert_code}
\end{figure}

Our covert channel utilizes contention on a functional unit, namely the \emph{floating point division unit} (see Figure~\ref{fig:backend}), to transmit data from transient instructions to non-transient instructions, which will retire and become architecturally visible. 
The floating point division unit was chosen as it is not fully pipelined (see Section~\ref{sec:whypipeline}) in all Intel, AMD, and ARM microarchitectures we tested. Table~\ref{tbl:micro_params} shows the tested microarchitectures, and their latency and throughput characteristics of the DIVSD instruction, which are obtained from~\cite{func_unit_timing_2019_abel}~\footnote{As defined in~\cite{func_unit_timing_2019_abel}, \emph{latency} refers to the clock cycles needed from the time the~$\upmu$op is issued to the time the result become available to dependent~$\upmu$ops, while \emph{throughput} refers to the clock cycles needed from the time the~$\upmu$op is issued until to the time the functional unit becomes available again.} Note that in all tested microarchitectures, the throughput of the \texttt{DIVSD} instruction is 4 or 8 cycles, meaning that while an DIVSD instruction is being executed, a pending DIVSD instruction has to wait 4 or 8 cycles before entering the floating point division unit. This delay makes the floating point division unit an ideal candidate for us to create a covert channel.

Figure~\ref{fig:func_covert_code} shows the code used to form the ideal covert channel. 
%The logical flow of the covert channel is the similar to that shown in figure~\ref{fig:attack_diagram}, but the complete flow is as follows: 
(1) A timer is started (Line 5);
(2) A chain of \emph{dependent} floating point division instructions begins execution (Line 8). Because the instructions are dependent, each instruction suffers the full round-trip latency of the floating point division unit (see Table~\ref{tbl:micro_params}). This chain of division instructions acts as a \emph{receiver};
(3) The result of the receiver instruction chain is compared in the if statement (Line 14). The if statement has been previously trained to be true, so the body will execute speculatively while the result of the receiver chain is being calculated;
(4) A single bit of the (secret) message to transmit is accessed (Line 15) and the inner if statement branches depending on the value of the secret bit (Line 16);
(5) The inner if statement is trained to be false. Thus, if the secret bit was `1',  the processor backtracks
%to the previous checkpoint (i.e., the inner if statement) 
and begins to speculatively execute a set of \emph{independent} floating point division instructions (Line 18-23), which act as a \emph{sender}.
The ``sender'' instructions are independent with each other so as to be issued concurrently and maximally contend with the ``receiver'' instructions on the floating point division unit of the processor. 
(6) When the ``receiver'' instructions are completed, the processor will realize the mis-speculation (\emph{recv} in Line 14 was 0) and squash the speculative instructions from the ``sender''. We then stop the timer (Line 28) and measure the time difference.

Note that if the secret bit was `1', the observed time difference will be longer, due to the contention in the floating point division unit with the mis-speculated ``sender'' instructions, compared to the case when the secret bit was `0' where there was no contention. 
This secret-dependent timing difference creates a covert channel.

\begin{table*} [htp]
\centering
\begin{tabular}{|c|c|c|c|c|c|}
  \hline
  \multirow{2}{*}{CPU} & \multirow{2}{*}{Microarch.} & Latency & Throughput & Transfer Rate & Error Rate \\
      &            & (cycles)& (cycles)   & (KB/s)        & (\%) \\
  \hline
  Intel Core i5-8250U     & Kabylake R   & 13--15 & 4 & 53.1 & 0.02\\  
  \hline
  Intel Core i5-6500     & Skylake   & 13--15 & 4 & 105.3 & <0.01\\
  \hline
  Intel Core i5-6200U     & Skylake   & 13--15 & 4 & 74.9 & 0.04\\  
%   Intel Core i5-8250U & Kaby Lake R & 13--15 & 4 & 2.347 & <0.01\\
%   % @ 1.60GHz   
%   \hline
%   Intel Xeon E5-2650 v4 & Broadwell & 10--14 & 4 & 1.697 & <0.01\\
%   % @ 2.20GHz
  \hline
  Intel Xeon E5-2658 v3 & Haswell & 10--20 & 8 & 64.1 & <0.01 \\
  % @ 2.20GHz  
  \hline
  Intel Core i5-3340M     & Ivybridge   & 10--20 & 8 & 75.6 & 0.16\\  
  \hline
  AMD Ryzen 3 2200G             & Zen       & 8--13 & 4 & 83.1 & 5.50\\ %1.163 & 0.003\\
  \hline
  AMD Ryzen 5 2600              & Zen+      & 8--13 & 4 & 84.8 & 3.30\\
  \hline
  NVIDIA Jetson Nano            & Cortex A57 & N/A & N/A & 87.7 & 0.02\\
  \hline
%   Raspberry Pi 4                & Cortex A72 & N/A & N/A & 80.7 & 0.16\\
%   \hline
  % Source: Cortex-A57 software optimization guide
  % FDIV/VDIV --> S-form: latency 7--17, throughput 2/15-2/5
  %               D-form: latency 7-32, throughput 1/30-1/5
\end{tabular}
\centering
\caption{Evaluation platforms; \texttt{DIVSD} (SSE floating point division) instruction latency and throughput~\cite{func_unit_timing_2019_abel} and performance (transfer and error rates) of the proposed floating point division unit based covert channel.}
\label{tbl:micro_params}
\end{table*}

\subsection{Covert Channel Properties}
% \subsection{Performance Analysis}

\begin{figure*}[htp]
    \centering
    % \begin{subfigure}[b]{0.45\textwidth}
    %     \includegraphics[width=\textwidth]{figs/kaby/kaby_turbo_off.pdf}
    %     \caption{Kaby Lake}
    %     \label{fig:time_kaby_turbo_off}
    % \end{subfigure}
%   \begin{subfigure}[b]{0.45\textwidth}
%        \includegraphics[width=\textwidth]{figs/kaby.pdf}
%        \caption{Kaby Lake Microarchitecture}
%        \label{fig:time_kaby}
%    \end{subfigure}
    % \begin{subfigure}[b]{0.45\textwidth}
    %     \includegraphics[width=\textwidth]{figs/broadwell.pdf}
    %     \caption{Broadwell}
    %     \label{fig:time_broadwell}
    % \end{subfigure}
    \begin{subfigure}[b]{0.45\textwidth}
        \includegraphics[width=\textwidth]{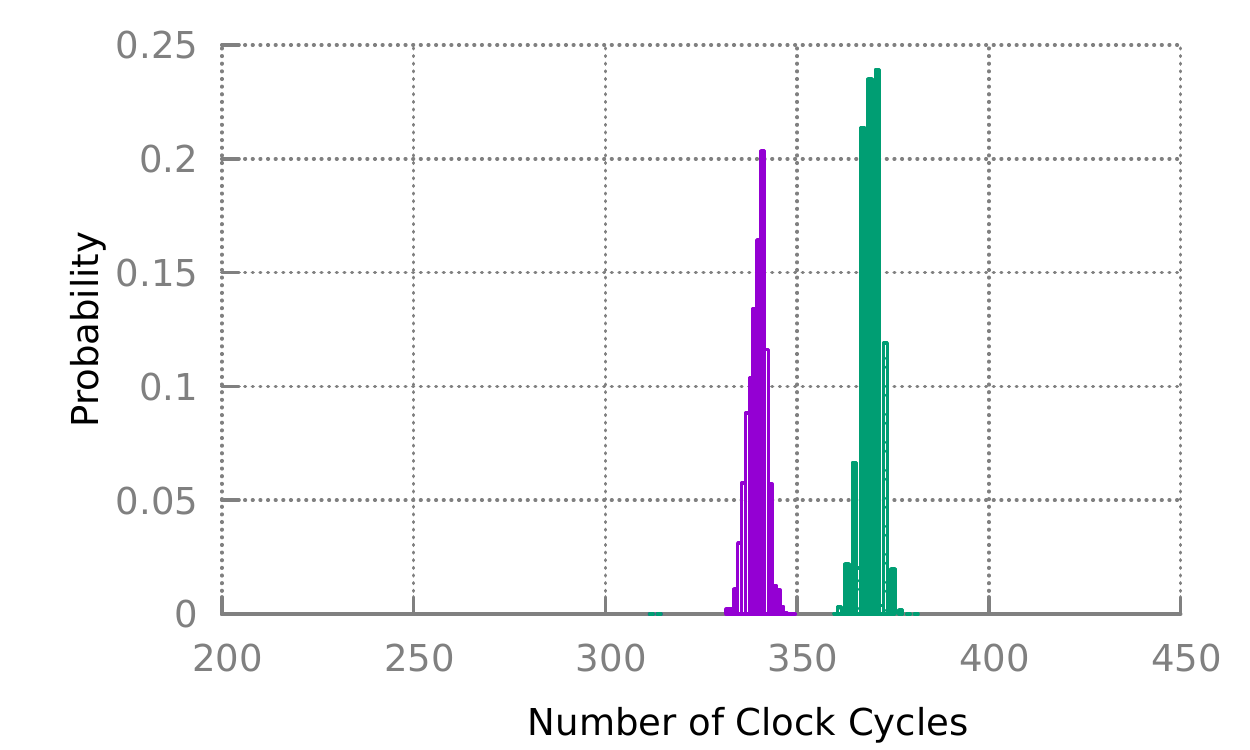}
        \caption{Kabylake R (i5-8250U)}
        \label{fig:time_sandybridge}
    \end{subfigure}
    \begin{subfigure}[b]{0.45\textwidth}
        \includegraphics[width=\textwidth]{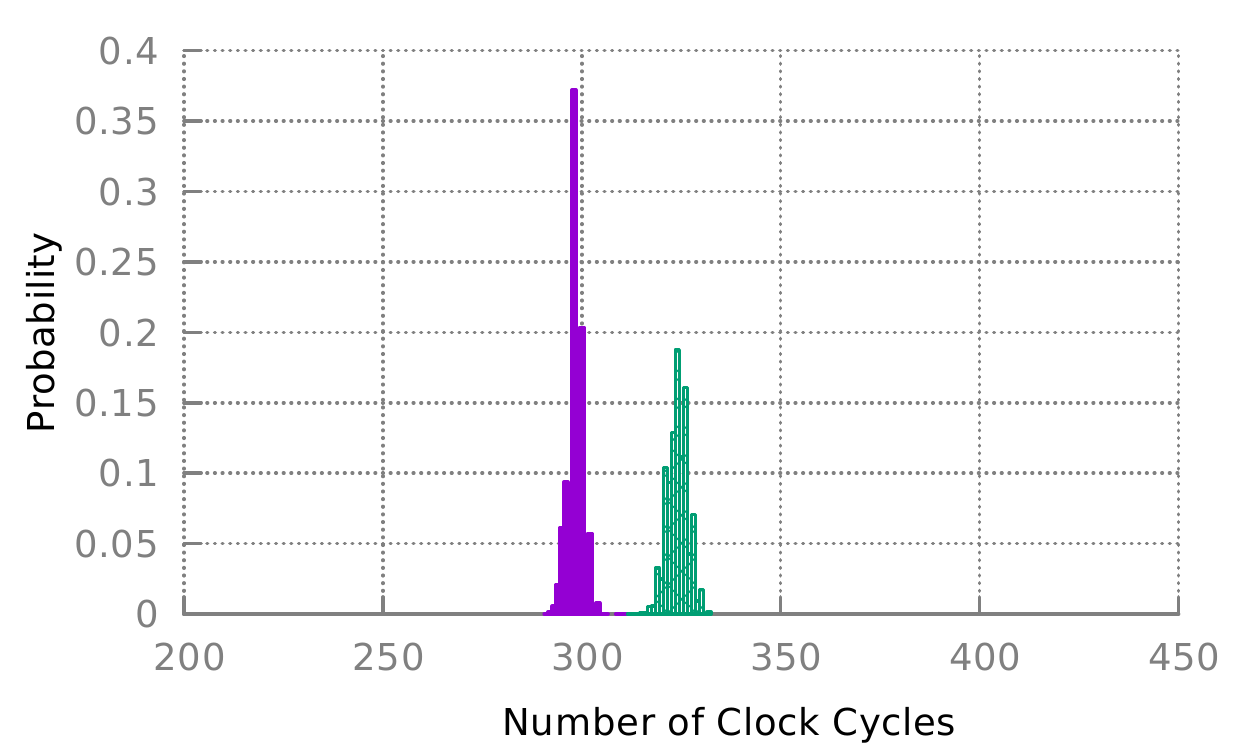}
        \caption{Skylake (i5-6500)}
        \label{fig:time_skylake_i5-6500}
    \end{subfigure}
    \begin{subfigure}[b]{0.45\textwidth}
        \includegraphics[width=\textwidth]{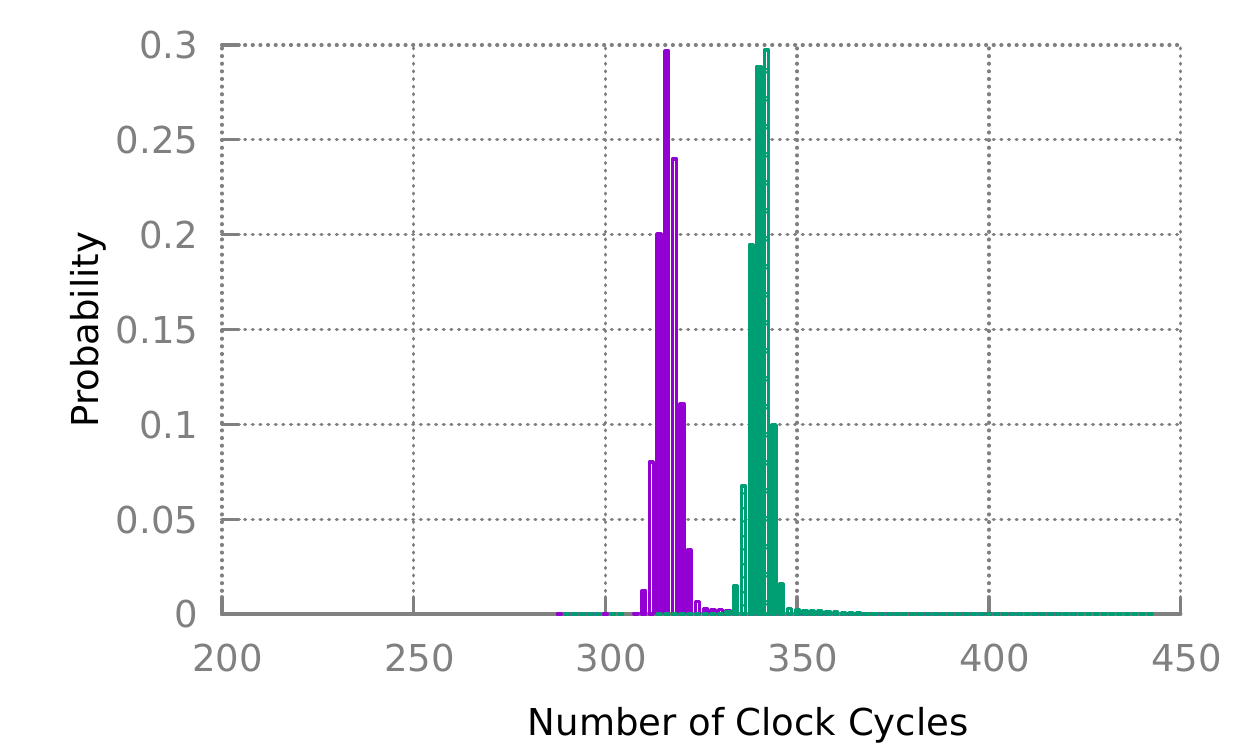}
        \caption{Skylake (i5-6200U)}
        \label{fig:time_skylake_i5-6200U}
    \end{subfigure}    
    \begin{subfigure}[b]{0.45\textwidth}
        \includegraphics[width=\textwidth]{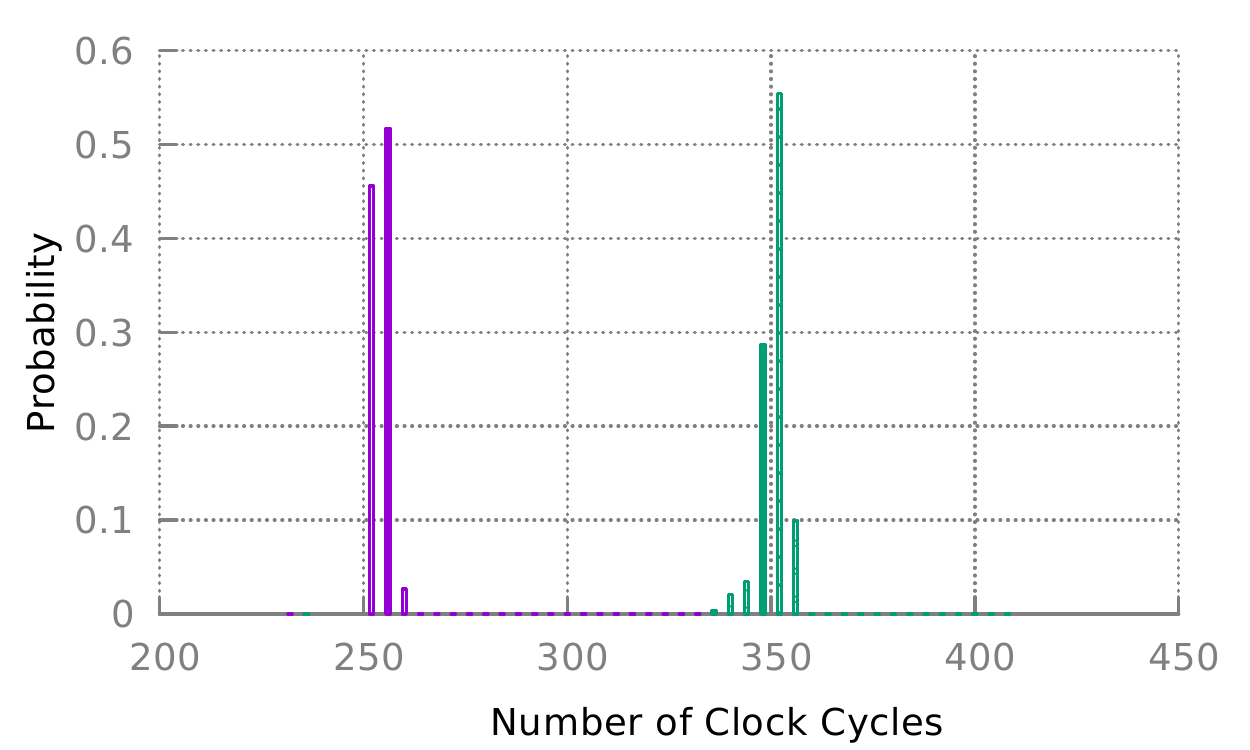}
        \caption{Haswell (E5-2658v3)}
        \label{fig:time_haswell}
    \end{subfigure}
    \begin{subfigure}[b]{0.45\textwidth}
        \includegraphics[width=\textwidth]{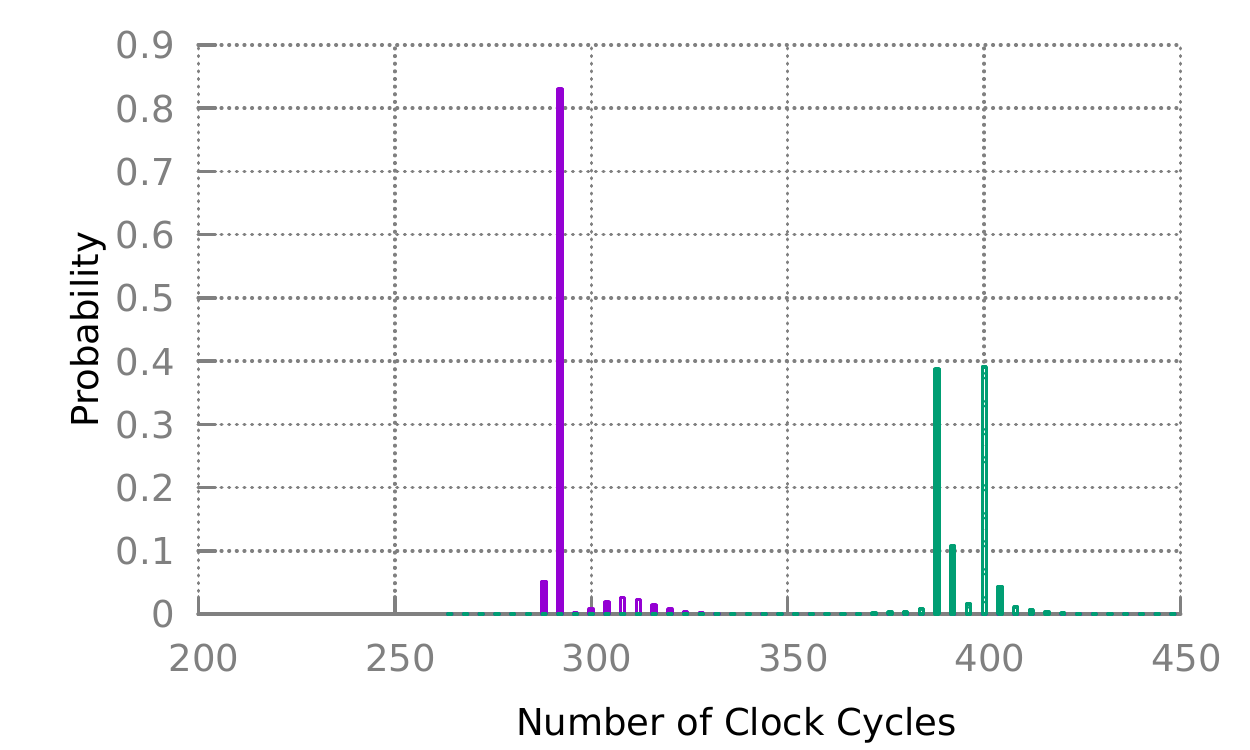}
        \caption{Ivybridge (i5-3340M)}
        \label{fig:time_ivybridge}
    \end{subfigure}    
    \begin{subfigure}[b]{0.45\textwidth}
        \includegraphics[width=\textwidth]{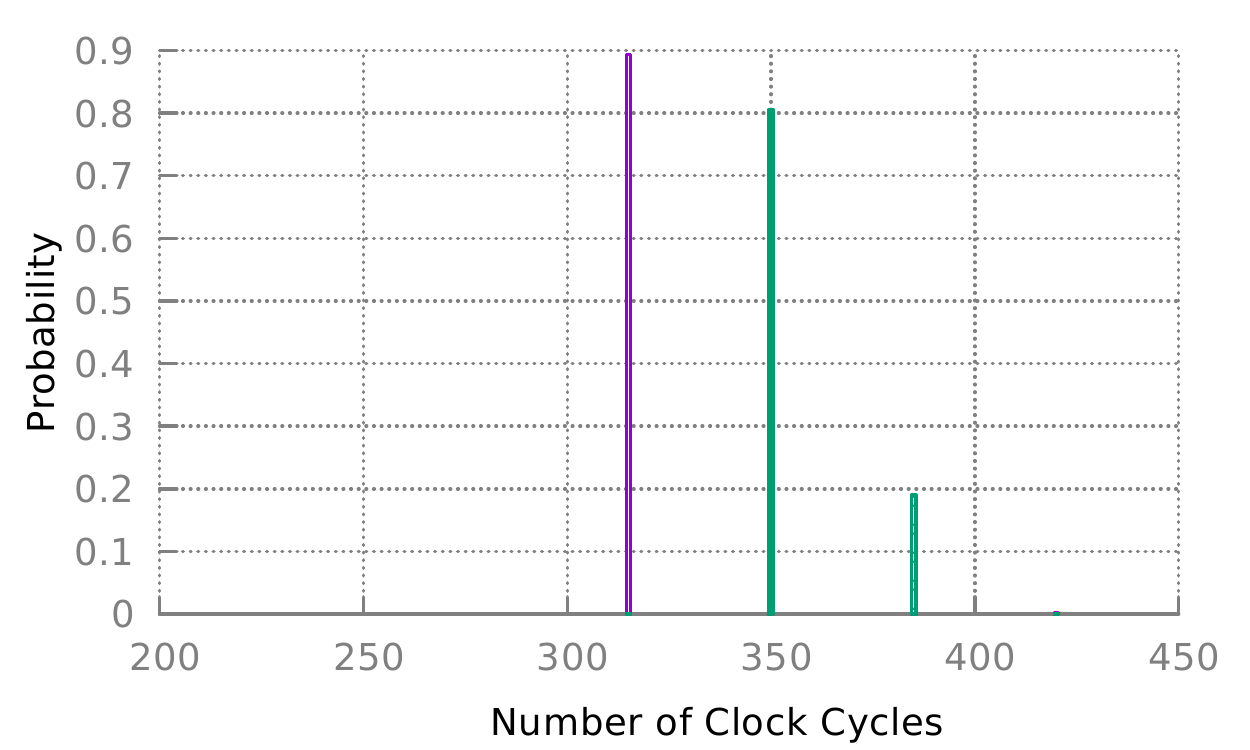} %{figs/zen.pdf}
        \caption{Zen (Ryzen3 2200G)}
        \label{fig:time_zen}
    \end{subfigure}
    \begin{subfigure}[b]{0.45\textwidth}
        \includegraphics[width=\textwidth]{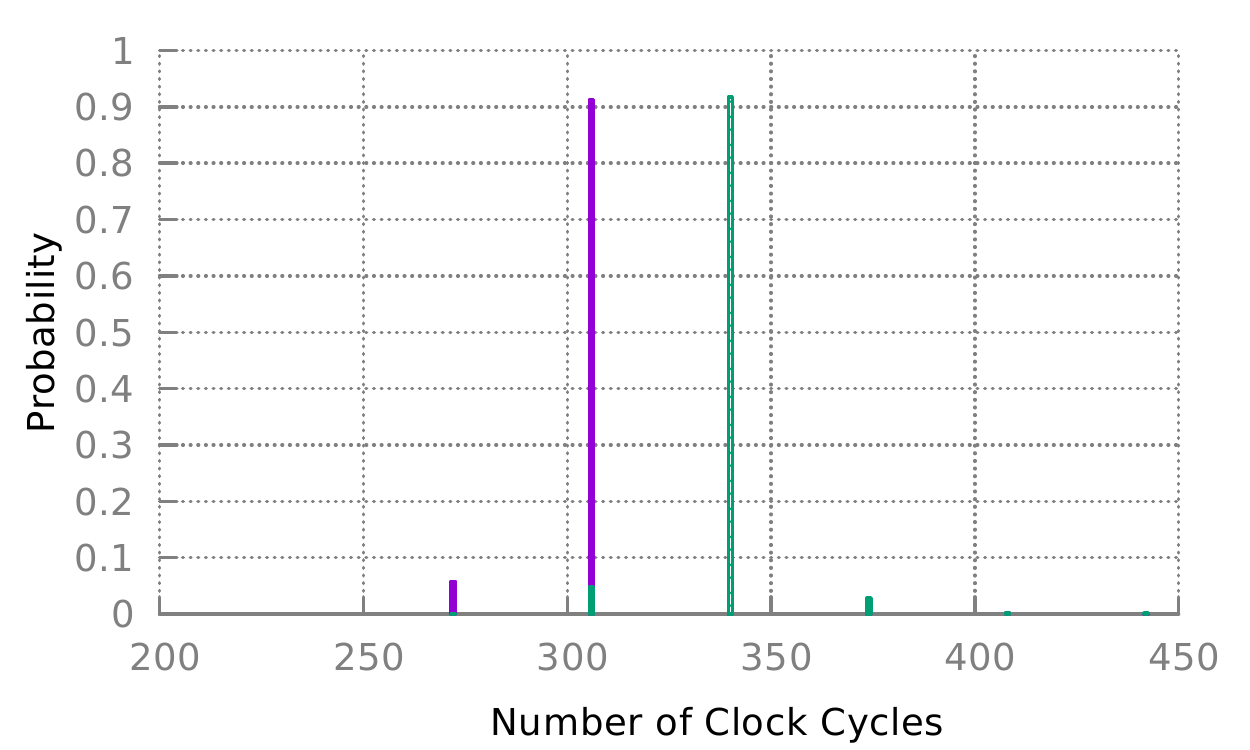} %{figs/zen.pdf}
        \caption{Zen+ (Ryzen5 2600)}
        \label{fig:time_zen+}
    \end{subfigure}
    \begin{subfigure}[b]{0.45\textwidth}
        \includegraphics[width=\textwidth]{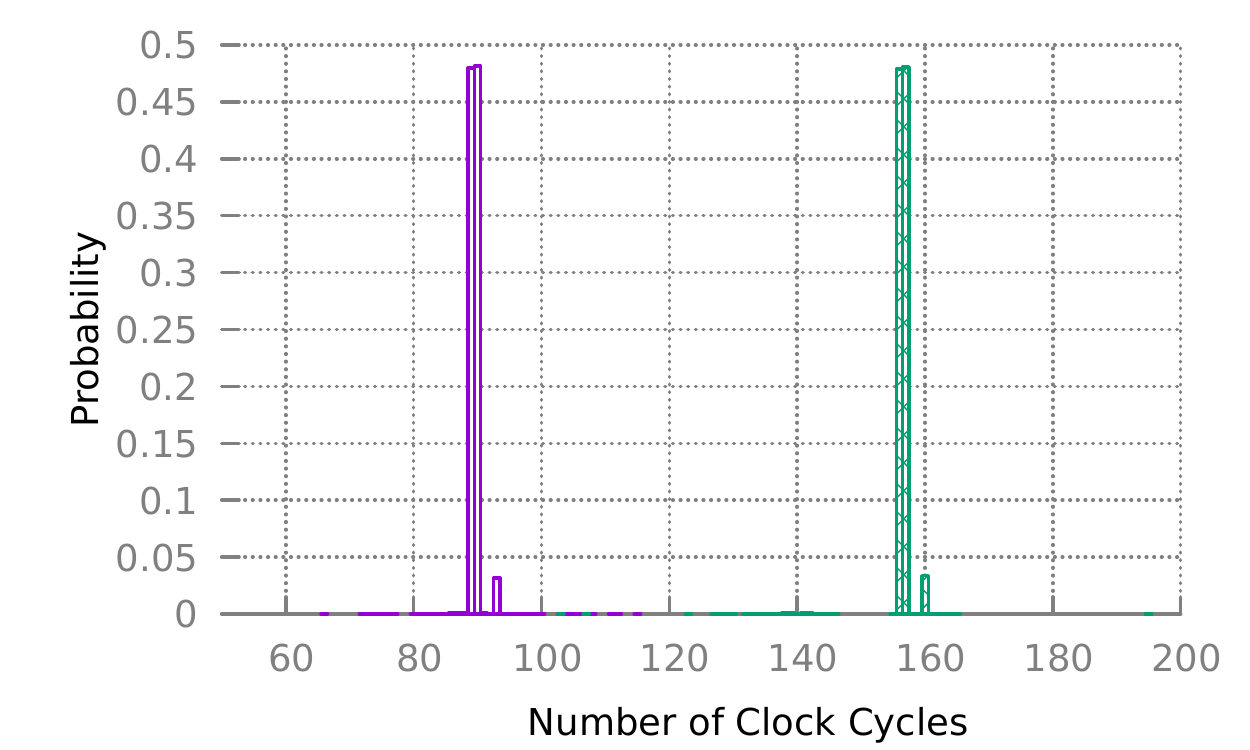} %{figs/zen.pdf}
        \caption{Cortex-A57 (Jetson Nano)*}
        \label{fig:time_zen+}
    \end{subfigure}    
    \caption{Floating point division unit based covert channel (Figure~\ref{fig:func_covert_code}) timing characteristics; 1,000,000 timing measurement samples of transmitting 0 (purple) and 1 (green). (*) For Cortex-A57, we use an additional thread based software counter for time measurement due to the lack of high-precision clock source (such as \texttt{rdtsc} in x86) available at the user level.}
    \label{fig:time_micro}
\end{figure*}

%\begin{figure}[htp]
%  \includegraphics[width=0.5\textwidth]{figs/kaby/kaby_turbo_on_snap.pdf}
%  \caption{This figuring shows snapshots of the kaby lake microarchitecture experiment. The snapshots are each 1000 samples wide. Here, we can see that while the latency of the floating point division unit changes overtime, we can still measure contention between the '0' and '1' values accurately.}
%  \label{fig:snapshot_kaby_turbo_boost}
%\end{figure}
%\fixme{just say 'turbo boost was disabled in Kaby Lake to reduce noise'}.

%\begin{figure}[htp]
%  \includegraphics[width=0.5\textwidth]{figs/kaby/kaby_turbo_warm.pdf}
%  \caption{This figure displays the properties of the warmup time for the kaby lake microarchitecture with turbo boost disabled. We witness a warmup period on the different microarchitecture for the different floating point division units.}
%  \label{fig:snapshot_kaby_turbo_boost}
%\end{figure}

% experiment setup
We experimentally evaluate the characteristics the covert channel on a number of commodity Intel, AMD, and ARM systems, as listed in Table~\ref{tbl:micro_params}.

Each system runs Linux (Ubuntu 18.04 or 16.04). For x86 platforms from Intel and AMD, we use \texttt{rdtscp} instructions for cycle accurate timing measurements. For ARM, we use an additional thread based software counter instead due to the architectural limitation. We repeatedly send 0 and 1 values over the covert channel, each for 1,000,000 times, and measure the timing results. To minimize noise, we use Linux's performance governor disable Turbo-boost (for X86 platforms) to improve reliability of the measurements. 

% experiment1: measure cycles to send 0 or 1. 
% result: transfer rate, error rate
Figure~\ref{fig:time_micro} shows the results. The X-axis shows the number of cycles taken to transmit, while the Y axis displays the probability a measurement has to take that many cycles.
Note first that on all tested platforms, we are able to see clear timing differences between `0' and `1' values. 
As explained in Section~\ref{sec:whypipeline}, not fully pipelined floating point division units in these platforms allow the mis-speculated division instructions to contend with the logically prior ``receiver'' instructions, resulting in clearly measurable timing differences. 

Another interesting observation is that the two AMD processors and the ARM Cortex-A57 show discreet timing characteristics---large proportion of the samples are concentrated on a few small measured cycles---whereas Intel processors show more varied timing behaviors, especially the Skylake processors. These differences are likely due to the way the floating point division unit is implemented in each of these vendors.

We also evaluated floating point multiplication instructions but were not able to observe any noticeable timing difference, suggesting that the floating point multiplication units in these platforms are well pipelined, and thus cannot be used to create covert channels. 

\subsection{Performance Analysis}
% experiment 2: measure transfer rate and error rate. 
% threshold calibration method, 
Next, we analyze the performance of the covert channel in terms of transfer rate and error rate. The measured transfer rates of our tested platforms are calculated by simply dividing the total bits sent (1 million bits of 0 and 1 million bits of 1) with the time it took to send them. 
The error rate of each system is calculated as follows. We first sort each million timing samples of 0 and 1. We then find 99 percentile value of the `0' samples and 1 percentile value of the `0' samples. If the former (99 percentile of `0' samples) is smaller than the latter (1 percentile of `1' samples), we pick the average of the two value as the threshold to determine 0 or 1. If the 99 percentile of 0 is bigger than the 1 percentile of 1, we set the average of the median values of 0 and 1 samples as the threshold value. We then apply the threshold against the collected samples to determine if it correctly classifies the sample against its known correct value.  

The results are shown in Table~\ref{tbl:micro_params} (see the `Transfer Rate' and `Error Rate' columns). First, notice that the proposed covert channel supports very high transfer rates on all tested platforms, ranging from 63 to 105 KB/s. Furthermore, the error rates are also very low, especially on Intel processors, as we observe less than 0.5\% error rates. AMD processors show higher error rates, of up to 5.5\% on low end Ryzen3 APU. 

\subsection{Sensitivity Analysis}
%\fixme{Need more details. need sensitivity analysis. \#of divs used. how long speculative window can be sustained? what's the maximum separation between 0 and 1? how to reduce noise? etc.}

An interesting aspect of our covert channel is that the size (duration) of the speculation window can be controlled by adjusting the number of dependent division instructions used in the ``receiver'' part of the covert channel---i.e., Line 8-11 in Figure~\ref{fig:func_covert_code}. This is because speculatively executed sender instructions are squashed after the receiver instruction change is completed. As such, the longer the receiver instruction chain is, the longer the sender instructions can contend on the floating point division unit. To understand the effect of the length of the receiver to the effectiveness of the covert channel, we measure the characteristics of the covert channel as a function of the number divisions in the receiver chain. 

\begin{table} [htp]
\centering
\begin{tabular}{|c|c|c|c|c|c|}
  \hline
  \multirow{2}{*}{\#divs} &  `0' &  `1' & Diff. & Transfer  & Error \\
                                 & (cycles) & (cycles) & (cycles) & (KB/s)       & (\%) \\
  \hline
 3 &	190	& 192	& 2	    & 102.9	& 45.77 \\
 6 &	232	& 236	& 4	    & 92.5	    & 11.95\\
 9 &	274	& 290	& 16	& 82.4	    & 0.70\\
12 &	314	& 342	& 28	& 74.9	    & 0.04\\
15 &	354	& 392	& 38	& 69.6	    & 0.18\\
24 &	394	& 436	& 42	& 55.7	    & 0.09\\
48 &	800	& 928	& 128	& 35.7	    & 0.43\\
72 &    1128 & 1252 & 137   & 27.3     & 0.12\\
% 80 &	1242 & 1376	& 134	& 25.2	    & 0.31\\
% 81 &	1230 & 1230	& 0	    & 25.1	    & 63.92\\
  \hline
\end{tabular}
\centering
\caption{Sensitivity to \#of divisions (\texttt{DIVSD}) used in the ``receiver" part of the covert channel on Intel Core i5-6200U.}
\label{tbl:sensitivity}
\end{table}

Table~\ref{tbl:sensitivity} shows the results. The first column shows the number of division instructions in the receiver chain. The second and third columns show the median cycles observed when sending `0' and `1' values over the covert channel, respectively. The fourth column is the cycle difference between 0 and 1 samples. Finally, the fifth and the last columns show the transfer and error rates of the channel.

Note first that the transfer rate is inversely proportional to the number of divisions in the receiver, which is expected as the more divisions are used, the longer time is needed to execute them before squashing the speculation. As such, from the transfer rate perspective, using a smaller number of divisions in the receiver may be desirable. However, when the number of divisions is too small, as in the case of 3 divisions, the covert channel becomes ineffective as the error rate is too high(>45\% error rate). This is because the speculation window is not long enough for the sender instructions to be able to effectively contend with the receiver instructions on the floating point division unit. 

The error rate dramatically decreases as we increase the number of divisions in the receiver. At 9 or more divisions
%---but less than 80---
the covert channel shows very low error rate while showing gradually decreasing transfer rates. 
%Interestingly, at 81 divisions, the covert channel suddenly stops functioning as we observe no difference in 0 and 1 timings. This is likely because the processor's capacity to buffer incoming instructions is exhausted at this point and therefore the speculative sender instructions cannot be executed. 
For this platform, we can see using 12 divisions in the receiver chain is a ``sweet spot'' in the sense that it offers high enough performance and low noise. While different platforms may have different sweet spots, we nevertheless used the same 12 divisions in all platforms, unless noted otherwise, as it performed reasonably well in all of them. 

\section{Transient Execution Attacks }\label{sec:SideChannel}

%\fixme{Need to re-written to show how SpectreRewind based covert-channels can be used in various realistic attack scenarios: (1) native code to dump kernel memory (Meltdown); (2) sandbox environments (javascript and/or bBPF) to  }
In this section, we present transient execution attacks using the proposed SpectreRewind \texttt{DIVSD} covert channel.

\subsection{Meltdown Attack}

%In the first scenario that we will consider, the attacker has complete control to select the assembly instructions that are executed during the attack. This includes a malicious user on a system that can craft binaries that contains a transient execution attack, and then run that binary attempting to direct access secret either from kernel or belonging to other user process on same system. This scenario is consistent with Meltdown-type attacks. 

% meltdown PoC: (1) attacker reads kernel memory from userspace.
Meltdown attacks~\cite{meltdown_2018_lipp} allow transient instructions to access secrets belonging to other processes and security domains, including the OS and virtual machines. In this section we describe our modifications to such attacks to utilize SpectreRewind covert channel.

In the Meltdown attack, the attacker attempts to read from a memory address, such as a kernel virtual address. While architecturally such an access will generate an exception, speculatively the access can forward secret data to a dependent load instruction, which encodes the secret into a cache state change, before it can be squashed. 

In our modification, we simply surround the exception generating memory access with DIVSD sender and receiver instructions as shown in Figure~\ref{fig:func_covert_code}. In more detail, we base our implementation on the original Meltdown open-source repository~\footnotemark. We modified a single function \texttt{libkdump\_read(addr)} in \texttt{libkdump.c}, which reads a single byte from the given address (\texttt{addr}), to utilize our SpectreRewind DIVSD covert channel. The rest of the code and other settings are unchanged.

Note that because a Meltdown attack generates exceptions, it is necessary to suppress such exceptions. In the original PoC, either Intel's Transactional Synchronization Extension (TSX) or signal handling was used to suppress  exceptions. In our approach, however, an exception generating secret memory access can only occur in a mis-speculated transient execution, which will be squashed when the the receiver code has completed. Thus, we effectively suppress the exception without needing to use TSX or signal handling methods used in the original Meltdown PoC.

\begin{table} [htp]
\centering
\begin{tabular}{|c|r|r|}
  \hline
Method  & \# Reads &  Success (\%) \\
  \hline
Original (SigHandle) &	2197	& 97.78 \\
Original (TSX) &	217691 &	100.00 \\
SpectreRewind	& 193399	& 99.97 \\
  \hline
\end{tabular}
\caption[]{Performance of SpectreRewind covert channel (\texttt{DIVSD}) based Meltdown attack (Demo \#3: Reliability test of the original Meltdown PoC repository) on Intel i5-6500.}
\label{tbl:meltdown_performance}
\end{table}
\footnotetext{\url{https://github.com/IAIK/meltdown}}

Table~\ref{tbl:meltdown_performance} compares the performance of our modified Meltdown attack  with the original ones, which utilize flush+reload based covert channels. Of the two original versions we evaluated, Original (SigHandle) suppresses exception by installing a signal handler while the Original (TSX) does so by utilizing TSX. We use the \emph{reliability} PoC in the official Meltdown repository, which continuously reads a single byte from a kernel memory address and reports the number of reads and the success (i.e., correct reading) rate. In each configuration, we ran the reliability PoC for 60 seconds and measured the performance on the Intel Core i5-6500 (Skylake) processor, which supports TSX. As can be seen in the table, our SpectreRewind version of Meltdown performs significantly faster than the signal handler version of the original Meltdown in terms of the speed and the success rate, while it performs similarly compared to the TSX version of the original attack.

\subsection{Sandbox (JavaScript) Attack}

\begin{figure}[h]
\centering
  \includegraphics[width=0.45\textwidth]{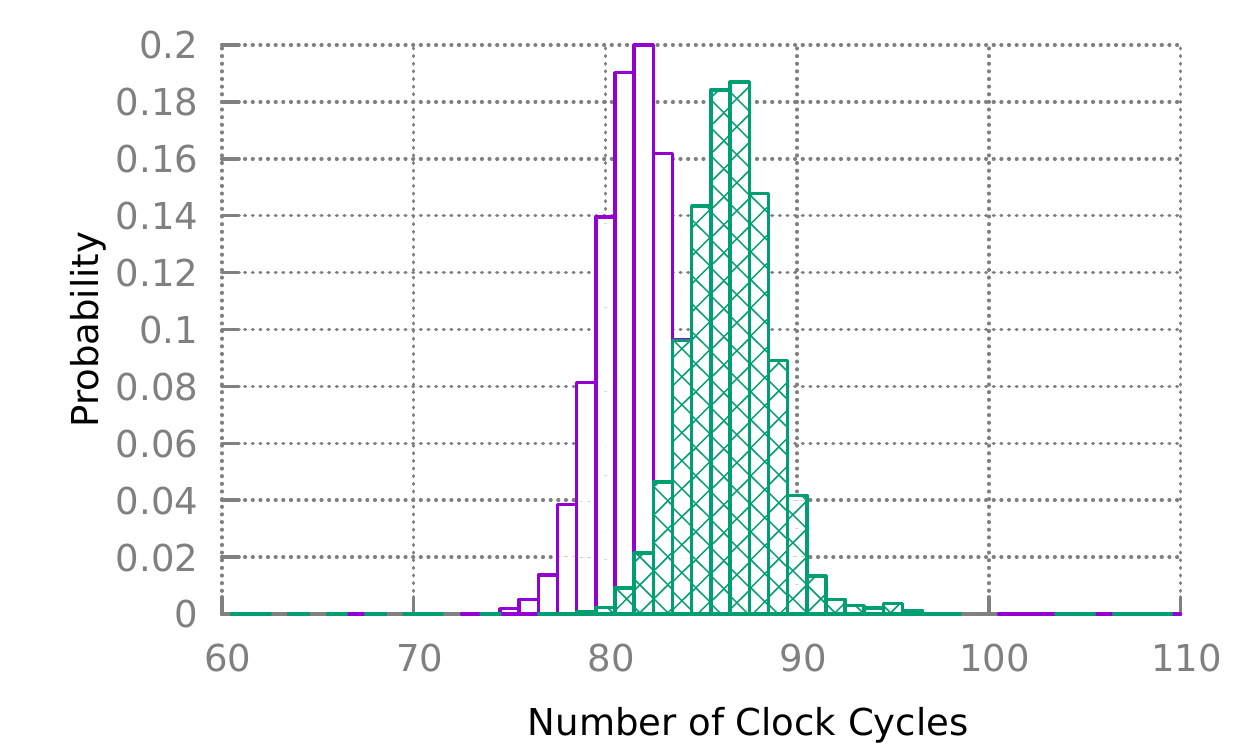}
  \caption{\label{fig:java_hist} Timing characteristics of division floating point unit covert channel execution in Google Chrome JavaScript sandbox}
\end{figure}

\begin{figure*}
\centering
  \includegraphics[width=0.95\textwidth]{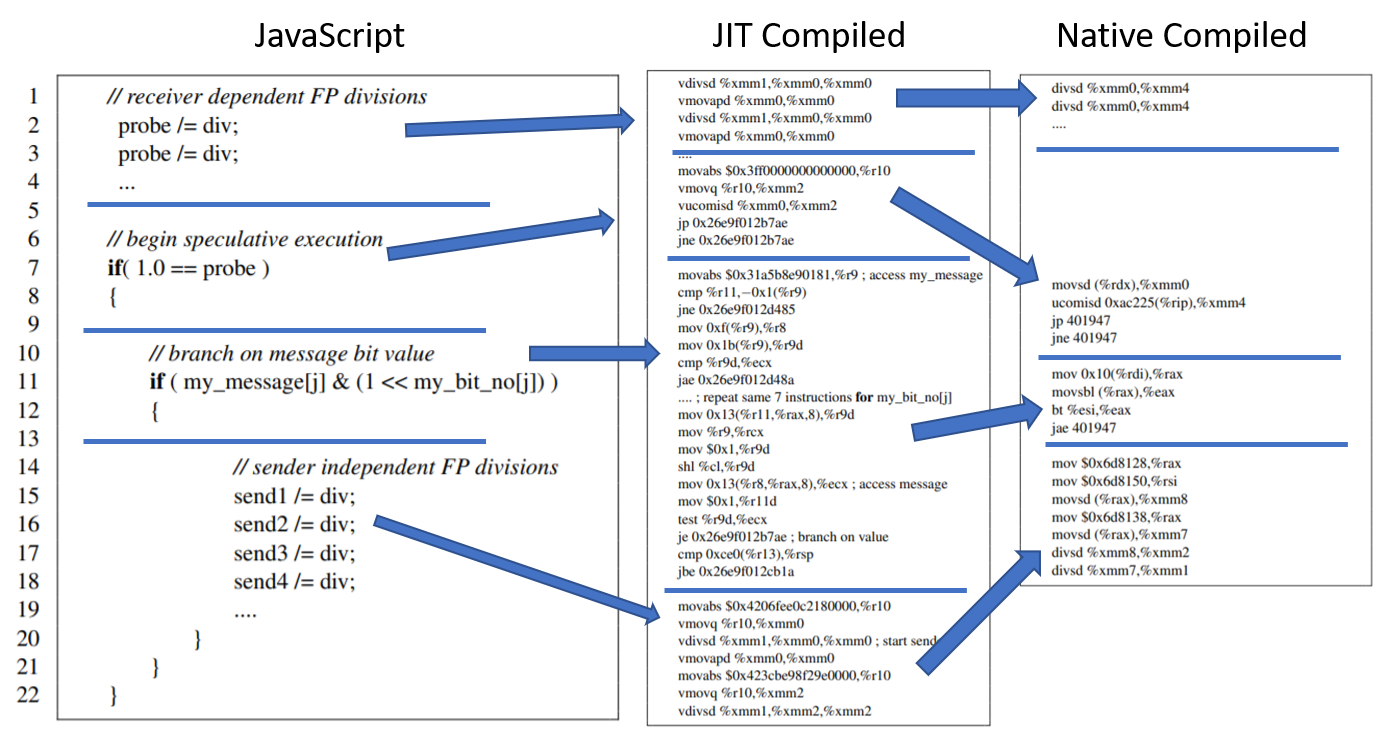}
  \caption{\label{fig:code_conv} Excerpt from JavaScript covert channel code (Left), the assembly the JIT compiler created (Center), and the native generated assembly (Right)}
\end{figure*}

One of the drawbacks of SpectreRewind is the amount of $\upmu$ops it requires to exist in the ReOrder Buffer simultaneously. While we have shown that this is not a problem when executing attacks in a native environment, sandbox environments---e.g. JavaScript---pose additional challenges. In this section, we show that it is possible to mount SpectreRewind attacks from sandbox environments by porting our floating point \texttt{DIVSD} based covert channel code to JavaScript, and successfully executing it on Google Chrome version 62.0.3202.75. For a high resolution timer, we take the same approach as past research~\cite{attacks_timers_fantastic} and utilize Web Workers along with \texttt{SharedArrayBuffer}. This allows for the creation of a separate thread, that continuously increments a value in shared memory that the original thread can use to time code execution. Overall, we found that minimal changes were needed to port our code. 

Figure~\ref{fig:code_conv} shows a snippet of the final code along with the generated assembly produced by the JavaScript JIT compiler side-by-side with the original attack code assembly. While the $\upmu$op footprint is increased from the native compiled version, we interestingly find that the majority of these extra instructions happen in the section of code that is responsible for accessing the message and branching on bit values. We see that the division operations are compiled down neatly into only a couple of floating point instructions, and we note that the extra vmovapd instructions added in the JIT compiler version do not take up any $\upmu$ops and thus the two areas of the code provide equivalent pressure on the ROB and scheduler. We also find the attack can still fit within the scheduler and ReOrder Buffer of a Kabylake R micro-architecture, and while the resolution of the SharedArrayBuffer is poor compared to native timers that the difference between timing bit values is sufficient for data transmission. We have however increased the number of receiver code divisions from 12 to 24 to improve signal over the lower resolution timer. We show the probability distribution of the transmission in Figure~\ref{fig:java_hist}.

\section{Discussion}\label{sec:eval}

In this section, we discuss the benefits and shortcomings of SpectreRewind, and its mitigation options.

% we also discuss both the proposed hardware defenses that it is able to circumvent, and proposed defenses that are immune to it.
% SpectreRewind is not a new transient execution attack. It is only a framework that allows the attacker to create covert channels that transmit the secret before the transient execution has been completed. This technique allows for the creation of a covert channel that uses contention on shared functional units during transient execution attacks. 
\subsection{Comparison to Cache Covert Channels}
To date, most transient execution attacks leverage cache based covert channels, especially Flush+Reload~\cite{related_flush_reload}, due to their high performance and low noise.
In this paper, we find that our floating point division unit based covert channel is available in a wide range of micro-architectures while providing similarly high performance and low noise characteristics. 
As such, we believe that our floating point division unit covert channel can be used as an alternative to cache based ones for transient execution attacks.
Our covert channel may be preferable to Flush+Reload in environments where instructions to flush cache lines (e.g. \texttt{CLFLUSH} in x86) are unavailable. In these environments, eviction sets must be created, but these channels can be noisy, especially in processors that use pseudo random cache replacement schemes~\cite{ARMageddon}. 

We also note that recently many researchers have proposed solutions to protect cache based covert channels. For example,  
InvisiSpec~\cite{solution_invisispec_2018_yan2018} and SafeSpec~\cite{solution_safespec_2018_khasawneh} are both recently proposed hardware solutions that defer updating microarchitectural states of caches (and TLBs) until such changes are considered to be safe. 
% Until that point, microarchitectural changes are instead stored within additional hardware structures, known as ``shadow buffers'', and are simply discarded if the instruction that would cause the change is squashed. 
Gonzalez et al~\cite{solution_RISCV_2019_Gonzalez} actually implemented such a defense on an out-of-order open source RISC-V processor core. 
CleanupSpec~\cite{solution_CleanupSpec_2019_Saileshwar} lets the microarchitectural changes from transient instructions to occur but later undo those changes after recognizing mis-speculation. 
SpectreRewind can bypass these defense mechanisms, as complete transmission of secret over the covert channel is accomplished---in forms of increased execution time of the receiver---before the transient instructions are completed, which make the aforementioned defense mechanisms ineffective.

% We also note that current research~\cite{scatterCache} suggests that high performance systems can be created that randomize cache sets, making it impractical to create eviction sets, so future micro-architectures may not have these well known cache options available. 

One downside of our approach is that it requires that sender and receiver instructions be present simultaneously at the same hardware thread, which restricts its use in cross-process attack scenarios (e.g., cross-process branch target injection attack~\cite{spectre_v1_v2_2019_Kocher}). 
In addition, the sender and receiver instructions must be in the same protection domain---either both in kernel or both in user. Therefore, initiating the receiver instructions at the user-level while executing the sender instructions at the kernel (e.g., a system call) may not be feasible.

\subsection{Mitigation Strategies}
% We now provide several mitigation strategies to mitigate SpectreRewind type covert channels.

% Mitigating against SpectreRewind using simultaneous covert channel. 
As SpectreRewind requires out-of-order contention on not fully pipelined functional units in the processor, one mitigation strategy is to redesign the functional units to be fully pipelined. But such a re-desing may not always be possible.  
% Covert channels utilizing concurrent contention functional units, like the floating point division unit, which we introduced in Section~\ref{sec:CovertChannel}, can be mitigated by fully pipelining the considered functional unit.
% Such a strategy would need to be employed in buffers and non-pipelined units within the limited resources of the memory system as well (e.g. MSHRs). 
Another alternative is to adopt a strict in-order scheduling policy such that younger instructions (sender) can never be issued before all older instructions are issued first, though it would incur high performance cost. 

% Functional units that cannot be redesigned in a fully pipelined fashion could also implement a scheduler that would block instructions bound for the functional unit until all other instructions bound for the same unit had been issued, basically making those functional units in-order while maintaining the out-of-order nature of the rest of the processor. 
% Finally, functional units could be made preemptive, allowing older instructions to preempt younger instructions that are currently using the units. The scheduler would need to be informed of this, and be redesigned to re-issue these instructions when the unit became available again.
% However, such changes may not always be possible due to various design constraints. 

SpectreRewind also requires secret data be forwarded transiently to the dependent instructions that cause the contention. Therefore, it can be mitigated through solutions that block or delay such forwarding. SpectreGuard~\cite{fustos2019spectreguard} is an example of such an approach, where secret data is marked as secret in the page tables and then is disallowed from being forwarded to dependent instructions until it reaches a point where it can be logically considered safe to forward. ConTExT~\cite{2019context} uses a similar approach, still marking data as secret in page tables and delaying forwarding of the secret value, but unlike SpectreGuard, not considering operations as safe until they reach the head of the ROB. Intel and NVIDIA also proposed new memory type based solutions~\cite{intel2019sapm, nvidia_patent}. STT~\cite{yu2019stt} improved upon these approaches, by considering if the instructions being forwarded secret data could transiently leak data into a covert channel, and if it could not, allowing that instruction to execute speculatively. 
All these techniques that prevent secret dependent speculative execution may mitigate SpectreRewind covert channels. 
\section{Related Work}\label{sec:related}

Cache based covert channels generally utilize the property that timing differences occur when accessing different levels of the cache, and these timing difference can be large---e.g. 1s of nanoseconds to access the L1 cache vs 100s of nanoseconds to access main memory on common high performance systems. Changes to the cache can also be long lasting (remain across context switches) as generally cache state changes remain until they are replaced by other memory accesses. Prime+Probe~\cite{related_prime_probe} takes advantage of the fact that caches are generally partitioned into sets that can hold a limited number of cache lines. Once the attacker finds a grouping that perfectly fills the set, measuring the access time of accessing the entire set allows the attacker to monitor other memory access across the system that are utilizing that same set which can be used as a side-channel to spy on a victim. Flush+Reload~\cite{related_flush_reload} is a cache based technique that uses special hardware instructions provided by architectures to flush cache lines from the caches. If the attacker shares memory with the victim, they need only flush a shared memory location, let the victim execute, and then time a reload to the memory location to determine if it was accessed by the attacker. Both techniques are commonly used in transient execution attacks to monitor cache activity performed by the transient instructions.  

Systems that implement simultaneous multi-threading (SMT), are particularly open to the creation of covert channels, as the multiple threads that run on a single core can potential share and thus compete for the resources on the core.
Wang and Lee explored functional unit sharing in SMT processors to create a covert channel~\cite{wang_2006_covert}. In their work, they created a covert channel---on a Pentium 4 processor---by utilizing contention on the shared integer multiplication unit. Concurrently, Acıiçmez and Seifert utilized contention on the same Intel processor---again using the shared integer multiplication unit---to create a microarchitectural side channel~\cite{Aciicmez_2007_side}. Utilizing this channel, the attacker could spy on another process running a square and multiply cryptographic function that was running concurrently on a separate hardware thread on the same core. Our work differs from these papers, as we explore the unique challenges of both creating similar functional unit covert channels, but from a non-SMT context---single hardware thread---and utilizing such contention in Spectre Attacks.

In 2016, Fogh introduced a technique called Covert Shotgun~\cite{smt_contention_2016_Fogh} in-which two processes running on threads in the same SMT core run through an iterative set of instruction groupings and time the results to determine if those instructions can cause measurable contention on the shared resources. Recently, two works have concurrently implemented such an approach to test the viability of port contention as a covert channel between two such processes. This approach utilizes the fact that ports may only issue one instruction per cycle to their underlying functional units, thus if both processes attempt to issue instructions that require functional units on the same port, one of the processes will need to stall that clock cycle---causing a measurable timing difference. PortSmash~\cite{port_smash_2019_Aldaya} utilized port contention to create a microarchitectural side-channel to leak the secret key from a vulnerable version of OpenSSL. SmotherSpectre~\cite{port_smother_2019_bhattacharyya} utilized port contention as the covert channel in a Branch Target Injection (BTI~\cite{spectre_v1_v2_2019_Kocher}) Spectre attack. Using BTI allowed this attack to run attacker code to transiently access secret in the victim and then to execute specific instructions---dependent on secret value---that could be easily detected by the attacker's process. Our work differs from all three approaches as we focus on non-SMT contention channels which face unique challenges, and from the latter two as we focus on functional unit contention, and not port contention.

%While this paper focused on a proposed type of hardware defense that is vulnerable to SpectreRewind attacks, other defenses are available that are not vulnerable.
%Conditional Speculation~\cite{solution_Con_Spec_2019_Li} proposed to block secrets from being forwarded to instructions that could modify the state of the data-cache. As this only blocks the data-cache covert channel, it is vulnerable to our covert channel

\section{Conclusion and Future Work}
In this paper, we showed that it is possible to create a covert channel utilizing concurrent contention on functional units from a single hardware thread. We introduced a new covert channel, which utilizes contention on the floating point division unit in commodity Intel, AMD, and ARM processors. Our covert channel achieved high performance and low noise comparable to that of the widely used Flush+Reload cache covert channel. 
We then showed that how the covert channel can be used in transient execution attacks. We implemented a Meltdown attack with our covert channel. We also showed that our covert channel can be used in the JavaScript sandbox of a Chrome browser. 
%While this covert channel does not effect the security of current software and hardware mitigation techniques, it does effect the security of proposed hardware solutions that may be implemented in the future and thus should be considered when designing such solutions.
% As future work, we would like to implement our framework within a sandbox environment, such as web browser's JavaScript engine, to test its viability across transient attacks on those environments. 
As future work, we plan to investigate if other microarchitectural structures can be used to create concurrent contention based covert channels. 

%\input{appendix}
%-----------------------------------------------------------------------
%%Acknowledgments
%\begin{acks}                            %% acks environment is optional
%This research is supported in part by NSF grant CNS 1718880 and NSA Science of Security initiative contract no. \#H98230-18-D-0009.
%\end{acks}

%\bibliographystyle{abbrv}
\bibliographystyle{ACM-Reference-Format}
%\bibliography{ccs-sample}
\bibliography{reference}


\begin{thebibliography}{00}

%%% the Web address.

\ifx \showCODEN    \undefined \def \showCODEN     #1{\unskip}     \fi
\ifx \showDOI      \undefined \def \showDOI       #1{#1}\fi
\ifx \showISBNx    \undefined \def \showISBNx     #1{\unskip}     \fi
\ifx \showISBNxiii \undefined \def \showISBNxiii  #1{\unskip}     \fi
\ifx \showISSN     \undefined \def \showISSN      #1{\unskip}     \fi
\ifx \showLCCN     \undefined \def \showLCCN      #1{\unskip}     \fi
\ifx \shownote     \undefined \def \shownote      #1{#1}          \fi
\ifx \showarticletitle \undefined \def \showarticletitle #1{#1}   \fi
\ifx \showURL      \undefined \def \showURL       {\relax}        \fi
% invisible to TeX
\providecommand\bibfield[2]{#2}
\providecommand\bibinfo[2]{#2}
\providecommand\natexlab[1]{#1}
\providecommand\showeprint[2][]{arXiv:#2}

\bibitem[\protect\citeauthoryear{??}{mel}{2018}]%
        {meltdown_3a_arm_2018}
 \bibinfo{year}{2018}\natexlab{}.
\newblock \showarticletitle{Cache Speculation Side-channels}.
\newblock \bibinfo{journal}{{\em ARM White paper\/}} (\bibinfo{year}{2018}).
\newblock


\bibitem[\protect\citeauthoryear{Abel and Reineke}{Abel and Reineke}{2019}]%
        {func_unit_timing_2019_abel}
\bibfield{author}{\bibinfo{person}{Andreas Abel} {and} \bibinfo{person}{Jan
  Reineke}.} \bibinfo{year}{2019}\natexlab{}.
\newblock \showarticletitle{uops.info: Characterizing Latency, Throughput, and
  Port Usage of Instructions on Intel Microarchitectures}. In
  \bibinfo{booktitle}{{\em Proceedings of the Twenty-Fourth International
  Conference on Architectural Support for Programming Languages and Operating
  Systems}} {\em (\bibinfo{series}{ASPLOS '19})}. \bibinfo{publisher}{ACM},
  \bibinfo{address}{New York, NY, USA}, \bibinfo{pages}{673--686}.
\newblock
\showISBNx{978-1-4503-6240-5}
\showDOI{%
\url{https://doi.org/10.1145/3297858.3304062}}


\bibitem[\protect\citeauthoryear{{Aciicmez} and {Seifert}}{{Aciicmez} and
  {Seifert}}{2007}]%
        {Aciicmez_2007_side}
\bibfield{author}{\bibinfo{person}{O. {Aciicmez}} {and} \bibinfo{person}{J.
  {Seifert}}.} \bibinfo{year}{2007}\natexlab{}.
\newblock \showarticletitle{Cheap Hardware Parallelism Implies Cheap Security}.
  In \bibinfo{booktitle}{{\em Workshop on Fault Diagnosis and Tolerance in
  Cryptography (FDTC 2007)}}. \bibinfo{pages}{80--91}.
\newblock
\showDOI{%
\url{https://doi.org/10.1109/FDTC.2007.16}}


\bibitem[\protect\citeauthoryear{Bhattacharyya, Sandulescu, Neugschwandtner,
  Sorniotti, Falsafi, Payer, and Kurmus}{Bhattacharyya et~al\mbox{.}}{2019}]%
        {port_smother_2019_bhattacharyya}
\bibfield{author}{\bibinfo{person}{Atri Bhattacharyya},
  \bibinfo{person}{Alexandra Sandulescu}, \bibinfo{person}{Matthias
  Neugschwandtner}, \bibinfo{person}{Alessandro Sorniotti},
  \bibinfo{person}{Babak Falsafi}, \bibinfo{person}{Mathias Payer}, {and}
  \bibinfo{person}{Anil Kurmus}.} \bibinfo{year}{2019}\natexlab{}.
\newblock \showarticletitle{SMoTherSpectre: exploiting speculative execution
  through port contention}.
\newblock \bibinfo{journal}{{\em arXiv preprint arXiv:1903.01843\/}}
  (\bibinfo{year}{2019}).
\newblock


\bibitem[\protect\citeauthoryear{Boggs, Segelken, Cornaby, Fortino, Chaudhry,
  Khartikov, Mooley, Tuck, and Vreugdenhil}{Boggs et~al\mbox{.}}{2019}]%
        {nvidia_patent}
\bibfield{author}{\bibinfo{person}{Darrell~D Boggs}, \bibinfo{person}{Ross
  Segelken}, \bibinfo{person}{Mike Cornaby}, \bibinfo{person}{Nick Fortino},
  \bibinfo{person}{Shailender Chaudhry}, \bibinfo{person}{Denis Khartikov},
  \bibinfo{person}{Alok Mooley}, \bibinfo{person}{Nathan Tuck}, {and}
  \bibinfo{person}{Gordon Vreugdenhil}.} \bibinfo{year}{2019}\natexlab{}.
\newblock \bibinfo{title}{Memory type which is cacheable yet inaccessible by
  speculative instructions}.
\newblock   (\bibinfo{date}{Jan.~3} \bibinfo{year}{2019}).
\newblock
\newblock
\shownote{US Patent App. 16/022,274.}


\bibitem[\protect\citeauthoryear{Cabrera~Aldaya, Bob~Brumley, ul~Hassan,
  Pereida~García, and Tuveri}{Cabrera~Aldaya et~al\mbox{.}}{2019}]%
        {port_smash_2019_Aldaya}
\bibfield{author}{\bibinfo{person}{Alejandro Cabrera~Aldaya},
  \bibinfo{person}{Billy Bob~Brumley}, \bibinfo{person}{Sohaib ul Hassan},
  \bibinfo{person}{Cesar Pereida~García}, {and} \bibinfo{person}{Nicola
  Tuveri}.} \bibinfo{year}{2019}\natexlab{}.
\newblock \showarticletitle{Port Contention for Fun and Profit}. In
  \bibinfo{booktitle}{{\em 2019 IEEE Symposium on Security and Privacy (SP)}}.
\newblock
\showDOI{%
\url{https://doi.org/10.1109/SP.2019.00066}}


\bibitem[\protect\citeauthoryear{Canella, Bulck, Schwarz, Lipp, von Berg,
  Ortner, Piessens, Evtyushkin, and Gruss}{Canella et~al\mbox{.}}{2018}]%
        {transient_systematic_2018_canella}
\bibfield{author}{\bibinfo{person}{Claudio Canella}, \bibinfo{person}{Jo~Van
  Bulck}, \bibinfo{person}{Michael Schwarz}, \bibinfo{person}{Moritz Lipp},
  \bibinfo{person}{Benjamin von Berg}, \bibinfo{person}{Philipp Ortner},
  \bibinfo{person}{Frank Piessens}, \bibinfo{person}{Dmitry Evtyushkin}, {and}
  \bibinfo{person}{Daniel Gruss}.} \bibinfo{year}{2018}\natexlab{}.
\newblock \showarticletitle{A Systematic Evaluation of Transient Execution
  Attacks and Defenses}.
\newblock \bibinfo{journal}{{\em CoRR\/}}  \bibinfo{volume}{abs/1811.05441}
  (\bibinfo{year}{2018}).
\newblock
\showeprint[arxiv]{1811.05441}
\showURL{%
\url{http://arxiv.org/abs/1811.05441}}


\bibitem[\protect\citeauthoryear{Fogh.}{Fogh.}{2016}]%
        {smt_contention_2016_Fogh}
\bibfield{author}{\bibinfo{person}{Anders Fogh.}}
  \bibinfo{year}{2016}\natexlab{}.
\newblock \showarticletitle{https://cyber.wtf/2016/09/27/covertshotgun/}.
\newblock  (\bibinfo{year}{2016}).
\newblock


\bibitem[\protect\citeauthoryear{Fustos, Farshchi, and Yun}{Fustos
  et~al\mbox{.}}{2019}]%
        {fustos2019spectreguard}
\bibfield{author}{\bibinfo{person}{Jacob Fustos}, \bibinfo{person}{Farzad
  Farshchi}, {and} \bibinfo{person}{Heechul Yun}.}
  \bibinfo{year}{2019}\natexlab{}.
\newblock \showarticletitle{{SpectreGuard: An Efficient Data-centric Defense
  Mechanism against Spectre Attacks}}. In \bibinfo{booktitle}{{\em DAC}}.
  \bibinfo{pages}{61--1}.
\newblock


\bibitem[\protect\citeauthoryear{Gonzalez, Korpan, Zhao, Younis, and
  Asanovi{\'c}}{Gonzalez et~al\mbox{.}}{2019}]%
        {solution_RISCV_2019_Gonzalez}
\bibfield{author}{\bibinfo{person}{Abraham Gonzalez}, \bibinfo{person}{Ben
  Korpan}, \bibinfo{person}{Jerry Zhao}, \bibinfo{person}{Ed Younis}, {and}
  \bibinfo{person}{Krste Asanovi{\'c}}.} \bibinfo{year}{2019}\natexlab{}.
\newblock \showarticletitle{Replicating and Mitigating Spectre Attacks on an
  Open Source RISC-V Microarchitecture}. In \bibinfo{booktitle}{{\em Third
  Workshop on Computer Architecture Research with RISC-V (CARRV)}}.
\newblock


\bibitem[\protect\citeauthoryear{Horn}{Horn}{2018}]%
        {spectre_v4_2018_horn}
\bibfield{author}{\bibinfo{person}{Jann Horn}.}
  \bibinfo{year}{2018}\natexlab{}.
\newblock \bibinfo{title}{speculative execution, variant 4: speculative store
  bypass}.
\newblock
  \bibinfo{howpublished}{\url{https://bugs.chromium.org/p/project-zero/issues/detail?id=1528}}.
    (\bibinfo{year}{2018}).
\newblock


\bibitem[\protect\citeauthoryear{Intel}{Intel}{2018}]%
        {meltdown_3a_intel_2018}
\bibfield{author}{\bibinfo{person}{Intel}.} \bibinfo{year}{2018}\natexlab{}.
\newblock \bibinfo{booktitle}{{\em {Intel Analysis of Speculative Execution
  Side Channels (Rev. 4.0)}}}.
\newblock \bibinfo{type}{{T}echnical {R}eport}.
\newblock
\showURL{%
\url{https://software.intel.com/sites/default/files/managed/b9/f9/336983-Intel-Analysis-of-Speculative-Execution-Side-Channels-White-Paper.pdf}}


\bibitem[\protect\citeauthoryear{Khasawneh, Koruyeh, Song, Evtyushkin,
  Ponomarev, and Abu-Ghazaleh}{Khasawneh et~al\mbox{.}}{2019}]%
        {solution_safespec_2018_khasawneh}
\bibfield{author}{\bibinfo{person}{Khaled~N. Khasawneh},
  \bibinfo{person}{Esmaeil~Mohammadian Koruyeh}, \bibinfo{person}{Chengyu
  Song}, \bibinfo{person}{Dmitry Evtyushkin}, \bibinfo{person}{Dmitry
  Ponomarev}, {and} \bibinfo{person}{Nael Abu-Ghazaleh}.}
  \bibinfo{year}{2019}\natexlab{}.
\newblock \showarticletitle{{SafeSpec: Banishing the Spectre of a Meltdown with
  Leakage-Free Speculation}}. In \bibinfo{booktitle}{{\em 56th Annual Design
  Automation Conference (ACM DAC)}}.
\newblock


\bibitem[\protect\citeauthoryear{Kiriansky and Waldspurger}{Kiriansky and
  Waldspurger}{2018}]%
        {spectre_v1.1_1.2_2018_kiriansky}
\bibfield{author}{\bibinfo{person}{Vladimir Kiriansky} {and}
  \bibinfo{person}{Carl Waldspurger}.} \bibinfo{year}{2018}\natexlab{}.
\newblock \showarticletitle{Speculative buffer overflows: Attacks and
  defenses}.
\newblock \bibinfo{journal}{{\em arXiv preprint arXiv:1807.03757\/}}
  (\bibinfo{year}{2018}).
\newblock


\bibitem[\protect\citeauthoryear{Kocher, Horn, Fogh, Genkin, Gruss, Haas,
  Hamburg, Lipp, Mangard, Prescher, Schwarz, and Yarom}{Kocher
  et~al\mbox{.}}{2019}]%
        {spectre_v1_v2_2019_Kocher}
\bibfield{author}{\bibinfo{person}{P. Kocher}, \bibinfo{person}{J. Horn},
  \bibinfo{person}{A. Fogh}, \bibinfo{person}{D. Genkin}, \bibinfo{person}{D.
  Gruss}, \bibinfo{person}{W. Haas}, \bibinfo{person}{M. Hamburg},
  \bibinfo{person}{M. Lipp}, \bibinfo{person}{S. Mangard}, \bibinfo{person}{T.
  Prescher}, \bibinfo{person}{M. Schwarz}, {and} \bibinfo{person}{Y. Yarom}.}
  \bibinfo{year}{2019}\natexlab{}.
\newblock \showarticletitle{Spectre Attacks: Exploiting Speculative Execution}.
  In \bibinfo{booktitle}{{\em 2019 IEEE Symposium on Security and Privacy
  (SP)}}. \bibinfo{publisher}{IEEE Computer Society}, \bibinfo{address}{Los
  Alamitos, CA, USA}.
\newblock
\showISSN{CFP19020-ART}
\showDOI{%
\url{https://doi.org/10.1109/SP.2019.00002}}


\bibitem[\protect\citeauthoryear{Koruyeh, Khasawneh, Song, and
  Abu-Ghazaleh}{Koruyeh et~al\mbox{.}}{2018}]%
        {spectre_ret_stack_2018_koruyeh}
\bibfield{author}{\bibinfo{person}{Esmaeil~Mohammadian Koruyeh},
  \bibinfo{person}{Khaled~N Khasawneh}, \bibinfo{person}{Chengyu Song}, {and}
  \bibinfo{person}{Nael Abu-Ghazaleh}.} \bibinfo{year}{2018}\natexlab{}.
\newblock \showarticletitle{Spectre returns! speculation attacks using the
  return stack buffer}. In \bibinfo{booktitle}{{\em WOOT}}.
\newblock


\bibitem[\protect\citeauthoryear{Lipp, Gruss, Spreitzer, Maurice, and
  Mangard}{Lipp et~al\mbox{.}}{2016}]%
        {ARMageddon}
\bibfield{author}{\bibinfo{person}{Moritz Lipp}, \bibinfo{person}{Daniel
  Gruss}, \bibinfo{person}{Raphael Spreitzer}, \bibinfo{person}{Cl{\'e}mentine
  Maurice}, {and} \bibinfo{person}{Stefan Mangard}.}
  \bibinfo{year}{2016}\natexlab{}.
\newblock \showarticletitle{ARMageddon: Cache Attacks on Mobile Devices}. In
  \bibinfo{booktitle}{{\em 25th {USENIX} Security Symposium ({USENIX} Security
  16)}}. \bibinfo{publisher}{{USENIX} Association}, \bibinfo{address}{Austin,
  TX}, \bibinfo{pages}{549--564}.
\newblock
\showISBNx{978-1-931971-32-4}
\showURL{%
\url{https://www.usenix.org/conference/usenixsecurity16/technical-sessions/presentation/lipp}}


\bibitem[\protect\citeauthoryear{Lipp, Schwarz, Gruss, Prescher, Haas, Fogh,
  Horn, Mangard, Kocher, Genkin, Yarom, and Hamburg}{Lipp
  et~al\mbox{.}}{2018}]%
        {meltdown_2018_lipp}
\bibfield{author}{\bibinfo{person}{Moritz Lipp}, \bibinfo{person}{Michael
  Schwarz}, \bibinfo{person}{Daniel Gruss}, \bibinfo{person}{Thomas Prescher},
  \bibinfo{person}{Werner Haas}, \bibinfo{person}{Anders Fogh},
  \bibinfo{person}{Jann Horn}, \bibinfo{person}{Stefan Mangard},
  \bibinfo{person}{Paul Kocher}, \bibinfo{person}{Daniel Genkin},
  \bibinfo{person}{Yuval Yarom}, {and} \bibinfo{person}{Mike Hamburg}.}
  \bibinfo{year}{2018}\natexlab{}.
\newblock \showarticletitle{Meltdown: Reading Kernel Memory from User Space}.
  In \bibinfo{booktitle}{{\em USENIX Security}}.
\newblock


\bibitem[\protect\citeauthoryear{Maisuradze and Rossow}{Maisuradze and
  Rossow}{2018}]%
        {spectre_ret2spec_2018_maisuradze}
\bibfield{author}{\bibinfo{person}{Giorgi Maisuradze} {and}
  \bibinfo{person}{Christian Rossow}.} \bibinfo{year}{2018}\natexlab{}.
\newblock \showarticletitle{{ret2spec: Speculative execution using return stack
  buffers}}. In \bibinfo{booktitle}{{\em ACM(CCS)}}. ACM,
  \bibinfo{pages}{2109--2122}.
\newblock


\bibitem[\protect\citeauthoryear{Minkin, Moghimi, Lipp, Schwarz, Van~Bulck,
  Genkin, Gruss, Sunar, Piessens, and Yarom}{Minkin et~al\mbox{.}}{2019}]%
        {mds_fallout_2019_Minkin}
\bibfield{author}{\bibinfo{person}{Marina Minkin}, \bibinfo{person}{Daniel
  Moghimi}, \bibinfo{person}{Moritz Lipp}, \bibinfo{person}{Michael Schwarz},
  \bibinfo{person}{Jo Van~Bulck}, \bibinfo{person}{Daniel Genkin},
  \bibinfo{person}{Daniel Gruss}, \bibinfo{person}{Berk Sunar},
  \bibinfo{person}{Frank Piessens}, {and} \bibinfo{person}{Yuval Yarom}.}
  \bibinfo{year}{2019}\natexlab{}.
\newblock \showarticletitle{{Fallout}: Reading Kernel Writes From User Space}.
\newblock


\bibitem[\protect\citeauthoryear{Oberman}{Oberman}{1999}]%
        {oberman1999floating}
\bibfield{author}{\bibinfo{person}{Stuart~F Oberman}.}
  \bibinfo{year}{1999}\natexlab{}.
\newblock \showarticletitle{Floating point division and square root algorithms
  and implementation in the AMD-K7/sup TM/microprocessor}. In
  \bibinfo{booktitle}{{\em IEEE Symposium on Computer Arithmetic (Cat. No.
  99CB36336)}}. IEEE, \bibinfo{pages}{106--115}.
\newblock


\bibitem[\protect\citeauthoryear{Saileshwar and Qureshi}{Saileshwar and
  Qureshi}{2019}]%
        {solution_CleanupSpec_2019_Saileshwar}
\bibfield{author}{\bibinfo{person}{Gururaj Saileshwar} {and}
  \bibinfo{person}{Moinuddin~K. Qureshi}.} \bibinfo{year}{2019}\natexlab{}.
\newblock \showarticletitle{CleanupSpec: An “Undo” Approach to Safe
  Speculation}. In \bibinfo{booktitle}{{\em Proceedings of the 52nd Annual
  IEEE/ACM International Symposium on Microarchitecture}} {\em
  (\bibinfo{series}{MICRO ’52})}. \bibinfo{publisher}{Association for
  Computing Machinery}, \bibinfo{address}{New York, NY, USA},
  \bibinfo{pages}{73–86}.
\newblock
\showISBNx{9781450369381}
\showDOI{%
\url{https://doi.org/10.1145/3352460.3358314}}


\bibitem[\protect\citeauthoryear{Schwarz, Lipp, Canella, Schilling, Kargl, and
  Gru{\ss}}{Schwarz et~al\mbox{.}}{2020}]%
        {2019context}
\bibfield{author}{\bibinfo{person}{Michael Schwarz}, \bibinfo{person}{Moritz
  Lipp}, \bibinfo{person}{{Claudio Alberto} Canella}, \bibinfo{person}{Robert
  Schilling}, \bibinfo{person}{Florian Kargl}, {and} \bibinfo{person}{Daniel
  Gru{\ss}}.} \bibinfo{year}{2020}\natexlab{}.
\newblock \showarticletitle{ConTExT: A Generic Approach for Mitigating
  Spectre}. In \bibinfo{booktitle}{{\em Network and Distributed System Security
  Symposium 2020}}.
\newblock
\showDOI{%
\url{https://doi.org/10.14722/ndss.2020.24271}}


\bibitem[\protect\citeauthoryear{Schwarz, Lipp, Moghimi, Van~Bulck, Stecklina,
  Prescher, and Gruss}{Schwarz et~al\mbox{.}}{2019}]%
        {mds_zombie_2019_Schwarz}
\bibfield{author}{\bibinfo{person}{Michael Schwarz}, \bibinfo{person}{Moritz
  Lipp}, \bibinfo{person}{Daniel Moghimi}, \bibinfo{person}{Jo Van~Bulck},
  \bibinfo{person}{Julian Stecklina}, \bibinfo{person}{Thomas Prescher}, {and}
  \bibinfo{person}{Daniel Gruss}.} \bibinfo{year}{2019}\natexlab{}.
\newblock \showarticletitle{{ZombieLoad}: Cross-Privilege-Boundary Data
  Sampling}. In \bibinfo{booktitle}{{\em CCS}}.
\newblock


\bibitem[\protect\citeauthoryear{Schwarz, Maurice, Gruss, and Mangard}{Schwarz
  et~al\mbox{.}}{2017}]%
        {attacks_timers_fantastic}
\bibfield{author}{\bibinfo{person}{Michael Schwarz},
  \bibinfo{person}{Cl{\'e}mentine Maurice}, \bibinfo{person}{Daniel Gruss},
  {and} \bibinfo{person}{Stefan Mangard}.} \bibinfo{year}{2017}\natexlab{}.
\newblock \showarticletitle{Fantastic Timers and Where to Find Them:
  High-Resolution Microarchitectural Attacks in JavaScript}. In
  \bibinfo{booktitle}{{\em Financial Cryptography and Data Security}},
  \bibfield{editor}{\bibinfo{person}{Aggelos Kiayias}} (Ed.).
  \bibinfo{publisher}{Springer International Publishing},
  \bibinfo{address}{Cham}, \bibinfo{pages}{247--267}.
\newblock
\showISBNx{978-3-319-70972-7}


\bibitem[\protect\citeauthoryear{Stecklina and Prescher}{Stecklina and
  Prescher}{2018}]%
        {meltdown_lazy_fp_2018_stecklina}
\bibfield{author}{\bibinfo{person}{Julian Stecklina} {and}
  \bibinfo{person}{Thomas Prescher}.} \bibinfo{year}{2018}\natexlab{}.
\newblock \showarticletitle{LazyFP: Leaking FPU Register State using
  Microarchitectural Side-Channels}.
\newblock \bibinfo{journal}{{\em arXiv preprint arXiv:1806.07480\/}}
  (\bibinfo{year}{2018}).
\newblock


\bibitem[\protect\citeauthoryear{Sun, Branco, and Hu}{Sun
  et~al\mbox{.}}{2019}]%
        {intel2019sapm}
\bibfield{author}{\bibinfo{person}{K Sun}, \bibinfo{person}{R Branco}, {and}
  \bibinfo{person}{K Hu}.} \bibinfo{year}{2019}\natexlab{}.
\newblock \showarticletitle{A New Memory Type Against Speculative Side Channel
  Attacks}.
\newblock  (\bibinfo{year}{2019}).
\newblock


\bibitem[\protect\citeauthoryear{Tromer, Osvik, and Shamir}{Tromer
  et~al\mbox{.}}{2010}]%
        {related_prime_probe}
\bibfield{author}{\bibinfo{person}{Eran Tromer}, \bibinfo{person}{Dag~Arne
  Osvik}, {and} \bibinfo{person}{Adi Shamir}.} \bibinfo{year}{2010}\natexlab{}.
\newblock \showarticletitle{Efficient Cache Attacks on AES, and
  Countermeasures}.
\newblock \bibinfo{journal}{{\em J. Cryptology\/}}  \bibinfo{volume}{23}
  (\bibinfo{date}{07} \bibinfo{year}{2010}), \bibinfo{pages}{37--71}.
\newblock
\showDOI{%
\url{https://doi.org/10.1007/s00145-009-9049-y}}


\bibitem[\protect\citeauthoryear{Tullsen, Eggers, and Levy}{Tullsen
  et~al\mbox{.}}{1995}]%
        {Tullsen-1995-SMT}
\bibfield{author}{\bibinfo{person}{Dean~M. Tullsen}, \bibinfo{person}{Susan~J.
  Eggers}, {and} \bibinfo{person}{Henry~M. Levy}.}
  \bibinfo{year}{1995}\natexlab{}.
\newblock \showarticletitle{Simultaneous Multithreading: Maximizing On-chip
  Parallelism}. In \bibinfo{booktitle}{{\em Proceedings of the 22Nd Annual
  International Symposium on Computer Architecture}} {\em
  (\bibinfo{series}{ISCA '95})}. \bibinfo{publisher}{ACM},
  \bibinfo{address}{New York, NY, USA}, \bibinfo{pages}{392--403}.
\newblock
\showISBNx{0-89791-698-0}
\showDOI{%
\url{https://doi.org/10.1145/223982.224449}}


\bibitem[\protect\citeauthoryear{Van~Bulck, Minkin, Weisse, Genkin, Kasikci,
  Piessens, Silberstein, Wenisch, Yarom, and Strackx}{Van~Bulck
  et~al\mbox{.}}{2018}]%
        {meltdown_foreshadow_2018_vanbulck}
\bibfield{author}{\bibinfo{person}{Jo Van~Bulck}, \bibinfo{person}{Marina
  Minkin}, \bibinfo{person}{Ofir Weisse}, \bibinfo{person}{Daniel Genkin},
  \bibinfo{person}{Baris Kasikci}, \bibinfo{person}{Frank Piessens},
  \bibinfo{person}{Mark Silberstein}, \bibinfo{person}{Thomas~F. Wenisch},
  \bibinfo{person}{Yuval Yarom}, {and} \bibinfo{person}{Raoul Strackx}.}
  \bibinfo{year}{2018}\natexlab{}.
\newblock \showarticletitle{Foreshadow: Extracting the Keys to the {Intel SGX}
  Kingdom with Transient Out-of-Order Execution}. In \bibinfo{booktitle}{{\em
  Proceedings of the 27th {USENIX} Security Symposium}}.
  \bibinfo{publisher}{{USENIX} Association}.
\newblock
\newblock
\shownote{See also technical report
  Foreshadow-NG~\cite{meltdown_foreshadow-NG_2018_weisse}.}


\bibitem[\protect\citeauthoryear{Van~Bulck, Moghimi, Schwarz, Lipp, Minkin,
  Genkin, Yuval, Sunar, Gruss, and Piessens}{Van~Bulck et~al\mbox{.}}{2020}]%
        {meltdown_LVI}
\bibfield{author}{\bibinfo{person}{Jo Van~Bulck}, \bibinfo{person}{Daniel
  Moghimi}, \bibinfo{person}{Michael Schwarz}, \bibinfo{person}{Moritz Lipp},
  \bibinfo{person}{Marina Minkin}, \bibinfo{person}{Daniel Genkin},
  \bibinfo{person}{Yarom Yuval}, \bibinfo{person}{Berk Sunar},
  \bibinfo{person}{Daniel Gruss}, {and} \bibinfo{person}{Frank Piessens}.}
  \bibinfo{year}{2020}\natexlab{}.
\newblock \showarticletitle{{LVI: Hijacking Transient Execution through
  Microarchitectural Load Value Injection}}. In \bibinfo{booktitle}{{\em 41th
  IEEE Symposium on Security and Privacy (S\&P'20)}}.
\newblock


\bibitem[\protect\citeauthoryear{van Schaik, Milburn, Österlund, Frigo,
  Maisuradze, Razavi, Bos, and Giuffrida}{van Schaik et~al\mbox{.}}{2019}]%
        {mds_ridl_2019_Schaik}
\bibfield{author}{\bibinfo{person}{Stephan van Schaik}, \bibinfo{person}{Alyssa
  Milburn}, \bibinfo{person}{Sebastian Österlund}, \bibinfo{person}{Pietro
  Frigo}, \bibinfo{person}{Giorgi Maisuradze}, \bibinfo{person}{Kaveh Razavi},
  \bibinfo{person}{Herbert Bos}, {and} \bibinfo{person}{Cristiano Giuffrida}.}
  \bibinfo{year}{2019}\natexlab{}.
\newblock \showarticletitle{{RIDL}: Rogue In-flight Data Load}. In
  \bibinfo{booktitle}{{\em S\&{P}}}.
\newblock


\bibitem[\protect\citeauthoryear{van Schaik, Minkin, Kwong, Genkin, and
  Yarom}{van Schaik et~al\mbox{.}}{2020}]%
        {mds_cacheOut}
\bibfield{author}{\bibinfo{person}{Stephan van Schaik}, \bibinfo{person}{Marina
  Minkin}, \bibinfo{person}{Andrew Kwong}, \bibinfo{person}{Daniel Genkin},
  {and} \bibinfo{person}{Yuval Yarom}.} \bibinfo{year}{2020}\natexlab{}.
\newblock \bibinfo{title}{{CacheOut}: Leaking Data on {Intel} {CPUs} via Cache
  Evictions}.
\newblock \bibinfo{howpublished}{\url{https://cacheoutattack.com/}}.
  (\bibinfo{year}{2020}).
\newblock


\bibitem[\protect\citeauthoryear{{Wang} and {Lee}}{{Wang} and {Lee}}{2006}]%
        {wang_2006_covert}
\bibfield{author}{\bibinfo{person}{Z. {Wang}} {and} \bibinfo{person}{R.~B.
  {Lee}}.} \bibinfo{year}{2006}\natexlab{}.
\newblock \showarticletitle{Covert and Side Channels Due to Processor
  Architecture}. In \bibinfo{booktitle}{{\em 2006 22nd Annual Computer Security
  Applications Conference (ACSAC'06)}}. \bibinfo{pages}{473--482}.
\newblock
\showISSN{1063-9527}
\showDOI{%
\url{https://doi.org/10.1109/ACSAC.2006.20}}


\bibitem[\protect\citeauthoryear{Weisse, Van~Bulck, Minkin, Genkin, Kasikci,
  Piessens, Silberstein, Strackx, Wenisch, and Yarom}{Weisse
  et~al\mbox{.}}{2018}]%
        {meltdown_foreshadow-NG_2018_weisse}
\bibfield{author}{\bibinfo{person}{Ofir Weisse}, \bibinfo{person}{Jo
  Van~Bulck}, \bibinfo{person}{Marina Minkin}, \bibinfo{person}{Daniel Genkin},
  \bibinfo{person}{Baris Kasikci}, \bibinfo{person}{Frank Piessens},
  \bibinfo{person}{Mark Silberstein}, \bibinfo{person}{Raoul Strackx},
  \bibinfo{person}{Thomas~F. Wenisch}, {and} \bibinfo{person}{Yuval Yarom}.}
  \bibinfo{year}{2018}\natexlab{}.
\newblock \showarticletitle{{Foreshadow-NG}: Breaking the Virtual Memory
  Abstraction with Transient Out-of-Order Execution}.
\newblock \bibinfo{journal}{{\em Technical report\/}} (\bibinfo{year}{2018}).
\newblock
\newblock
\shownote{See also {USENIX} Security paper
  Foreshadow~\cite{meltdown_foreshadow_2018_vanbulck}.}


\bibitem[\protect\citeauthoryear{Yan, Choi, Skarlatos, Morrison, Fletcher, and
  Torrellas}{Yan et~al\mbox{.}}{2018}]%
        {solution_invisispec_2018_yan2018}
\bibfield{author}{\bibinfo{person}{Mengjia Yan}, \bibinfo{person}{Jiho Choi},
  \bibinfo{person}{Dimitrios Skarlatos}, \bibinfo{person}{Adam Morrison},
  \bibinfo{person}{Christopher~W Fletcher}, {and} \bibinfo{person}{Josep
  Torrellas}.} \bibinfo{year}{2018}\natexlab{}.
\newblock \showarticletitle{{InvisiSpec: Making Speculative Execution Invisible
  in the Cache Hierarchy}}. In \bibinfo{booktitle}{{\em International Symposium
  on Microarchitecture (MICRO)}}.
\newblock


\bibitem[\protect\citeauthoryear{Yarom and Falkner}{Yarom and Falkner}{2014}]%
        {related_flush_reload}
\bibfield{author}{\bibinfo{person}{Yuval Yarom} {and} \bibinfo{person}{Katrina
  Falkner}.} \bibinfo{year}{2014}\natexlab{}.
\newblock \showarticletitle{FLUSH+RELOAD: A High Resolution, Low Noise, L3
  Cache Side-Channel Attack}. In \bibinfo{booktitle}{{\em 23rd {USENIX}
  Security Symposium ({USENIX} Security 14)}}. \bibinfo{publisher}{{USENIX}
  Association}, \bibinfo{address}{San Diego, CA}, \bibinfo{pages}{719--732}.
\newblock
\showISBNx{978-1-931971-15-7}
\showURL{%
\url{https://www.usenix.org/conference/usenixsecurity14/technical-sessions/presentation/yarom}}


\bibitem[\protect\citeauthoryear{Yu, Yan, Khyzha, Morrison, Torrellas, and
  Fletcher}{Yu et~al\mbox{.}}{2019}]%
        {yu2019stt}
\bibfield{author}{\bibinfo{person}{Jiyong Yu}, \bibinfo{person}{Mengjia Yan},
  \bibinfo{person}{Artem Khyzha}, \bibinfo{person}{Adam Morrison},
  \bibinfo{person}{Josep Torrellas}, {and} \bibinfo{person}{Christopher~W
  Fletcher}.} \bibinfo{year}{2019}\natexlab{}.
\newblock \showarticletitle{Speculative Taint Tracking (STT) A Comprehensive
  Protection for Speculatively Accessed Data}. In \bibinfo{booktitle}{{\em
  International Symposium on Microarchitecture (MICRO)}}.
  \bibinfo{pages}{954--968}.
\newblock


\end{thebibliography}
\end{document}